\documentclass[fleqn,usenatbib]{mnras}

\usepackage{newtxtext,newtxmath}
\usepackage[T1]{fontenc}
\usepackage{ae,aecompl}

\usepackage{graphicx}	% Including figure files
\usepackage{amsmath}	% Advanced maths commands
\usepackage{amssymb}	% Extra maths symbols
\usepackage{etoolbox}
\makeatletter
\patchcmd\@combinedblfloats{\box\@outputbox}{\unvbox\@outputbox}{}{\errmessage{\noexpand patch failed}}
\makeatother
\title[Wide binary companions and the weakest natal kicks]{Wide binary companions to massive stars \\ and their use in constraining natal kicks }

\author[A.P. Igoshev \& H.B. Perets]{
Andrei P. Igoshev,$^{1}$\thanks{E-mail: ignotur@gmail.com}
Hagai B. Perets,$^{1}$
\\
$^{1}$Physics Department, Technion - Israel institute of Technology, Haifa 3200002, Israel
}

\date{Accepted XXX. Received YYY; in original form ZZZ}

\pubyear{2015}

\begin{document}
\label{firstpage}
\pagerange{\pageref{firstpage}--\pageref{lastpage}}
\maketitle

\begin{abstract}
The origin of ultra-wide massive binaries (orbital separations $10^3-2\times 10^5$~AU) and their properties are not well characterized nor understood. Here we use the second Gaia data release to search for wide astrometric companions to Galactic O-B5 stars which share similar parallax and proper motion with the primaries. Using the data we characterize the frequency and properties of such binaries. We find an ultra-wide multiplicity fraction of $4.4\pm0.5$ per cent, to our completeness limit (up to $\approx 17$~mag; down to G-stars at distances of 0.3-2~kpc, excluding stars in clusters). The secondary mass-function is generally consistent with a Kroupa initial stellar function; if extrapolated to lower mass companion stars we then might expect a wide-binary fraction of $\sim 27\pm5\%$. In addition we use these data as a verification sample to test the existence of ultra-wide binaries among neutron stars (NSs) and black holes (BHs). We propose that the discovery of such binary can provide unique constraints on the weakest natal kicks possible for NSs/BHs. If a compact object is formed in an ultra-wide binary and receives a very-low natal kick, such a binary should  survive as a common proper motion pair. We therefore use Gaia data to search for ultra-wide companions to pulsars (normal and millisecond ones) and X-ray binaries. We find no reliable pairs. Future data could potentially provide stringent constraints through this method. 
\end{abstract}


%%%%%%%%%%%%%%%%%%%%%%%%%%%%%%%%%%%%%%%%%%%%%%%%%%

%%%%%%%%%%%%%%%%% BODY OF PAPER %%%%%%%%%%%%%%%%%%

\section{Introduction}
High-mass binaries have been extensively studied both observationally and theoretically (\citealt{sana2012,Duc+13,moe2017}). Close binaries play a critical role as the progenitors of type Ib/Ic supernovae and compact binaries, including X-ray binaries and gravitational-wave (GW) sources. Ultra-wide binaries ($>10^3$ AU), however, received less attention. 

Most stars are thought to form in stellar associations and clusters. In such dense environments ultra-wide binaries are "soft" \citep{1975MNRAS.173..729H}.  Even if they form in such environments, they are likely to be dynamically destroyed on short time-scales through binary-single and binary-binary interactions. The origins of wide binaries are therefore not well understood. Such soft binaries were suggested to be continuously destroyed and reform through dynamical captures, and those existing during the final stages and evaporation of the birth cluster/association, can then survive after the association/cluster dispersal \citep{2010MNRAS.404.1835K,2010MNRAS.404..721M,2011MNRAS.415.1179M} and give rise to observed field wide-binaries. In principal, wide binaries could later be destroyed due to flybys in the field, however, these processes are typically not important for most of the relatively young objects studied here (see Appendix~\ref{a:lifetime} for a brief discussion of this issue).

Although ultra-wide low-mass binaries have been explored and characterized \citep{1991A&A...248..485D,2010ApJS..190....1R}, their correspondent massive ultra-wide binaries, and in particular Galactic ones have been little explored. Here we use the new information from the second Gaia data release to identify and characterize Galactic massive (O-B5) main-sequence ultra-wide binaries ($10^3-2\times 10^5$  AU separations). In this article we use the definition of wide binary following \cite{2017MNRAS.472..675A} and assume that two (or more) stars form wide binary if  their properties are consistent with being gravitationally bound to each other on timescales comparable to the lifetime of the primary star. 

Since massive stars are the progenitors of neutron stars (NSs) and black-holes (BHs), we also search for ultra-wide binaries among such compact objects, including ultra-wide companions to pulsars and millisecond pulsars (MSPs) and X-ray binaries (both high-mass and low-mass binaries). Since mass loss and NS/BH natal kicks affect the evolution of such objects, ultra-wide binaries are not likely to be found among them. Nevertheless, as we discuss below, any discovery of ultra-wide bound companions to such objects could provide unique constraints on the existence and distribution of very low (or even zero) velocity natal kicks. We therefore provide a brief background on these issues before discussing our methods and results. 

The article is structured as following: we begin with the description of formation of compact objects and compact binaries  (Section~\ref{s:ns}), further we describe our method (Section~\ref{s:method}) to identify ultra-wide pairs. We then describe our sample selection and our analysis of the samples of O/B stars (Section~\ref{s:verif}), pulsars (Section~\ref{s:pulsar_sources}) and X-ray binaries (Section~\ref{s:xrb}). 
%In the last Section we discuss limitations of our search and the implications for the survival of wide binaries and then conclude.

\section{Ultra-wide BH and NS systems as probes for the weakest natal kicks}
%\section{Ultra-wide companions to NSs and BHs}
\label{s:ns}
A compact object (NS or BH) is born in a supernova explosion. The observed high velocities of NSs and some BH-binaries compared with the lower velocities typically observed for their progenitor OB massive stars suggest that both NSs and BHs are kicked at the moment of the explosion or immediately after. The origin of this velocity kick is still under debate \citep{2001LNP...578..424L}.  The inferred NS natal kicks velocities range from possibly zero km~s~$^{-1}$ to $\approx1000$~km~s$^{-1}$ with an average speed of 364~km~s$^{-1}$ \citep{verbunt2017, 1994Natur.369..127L}, while only $\sim$3 per cent of pulsars appear to be born with velocities smaller than 60~km~s$^{-1}$. The distribution of stellar-mass BH natal kicks is not known. The individual natal kicks for stellar-mass BH are usually estimated indirectly based on spatial distribution of low-mass X-ray binaries and seem to reach $\sim 100$~km~s$^{-1}$ \citep{2004MNRAS.354..355J,2012MNRAS.425.2799R,2017MNRAS.467..298R}. 

Some of the possible formation channels for NSs include cases where little mass loss and weak (or even zero) natal kicks are involved such as the case of an accretion-induced collapse of a white dwarf  \citep{1991ApJ...367L..19N} or an electron capture supernova explosion \citep{1980PASJ...32..303M, 1984ApJ...277..791N, 1987ApJ...322..206N}. The formation mechanism for BH natal kicks is also not understood.  Although in some scenarios both NS and BH kicks can arise from the same mechanism (e.g. in the gravitational tug-boat mechanism by  \citealt{2017ApJ...837...84J,2013MNRAS.434.1355J}), they might have very different origins and/or properties. The lowest possible value for a natal kick to BH or NS is not known. As a result, population synthesis studies need to make specific assumptions regarding the distribution, and in some of them it is assumed that the natal kick can be exactly zero for a significant fraction of the objects (see e.g. \citealt{2019MNRAS.482.2234G} for a recent double NS population synthesis and \citealt{2018A&A...619A..53T} for the white dwarf-NS merger rates). This assumption has to be tested observationally.  

An X-ray binary (XRB) contains a compact object: NS or BH and a secondary star (donor) which loses material. XRBs are usually classified as low-mass or high-mass X-ray binaries (LMXBs or HMXBs) if the secondary mass is below $2M_\odot$ or above $8M_\odot$, respectively.  An NS accreting gas in a LMXB can be spun up to form a millisecond pulsar (MSP) with a white dwarf companion at later stages of its evolution \citep{1991PhR...203....1B,tauris2011}.

If a fraction of all compact objects receive no natal kick and lose only a moderate amount of mass, a third/second component with a large orbital separation can survive the supernova explosion and continue orbiting the XRB, MSP or normal pulsar, respectively. In this work we focus on ultra-wide binaries with orbital separations of $10^3-2\times 10^5$~a.u. The corresponding large orbital periods of 0.1-10~Myr therefore prevent the detection of such companions by the timing technique in the case of normal pulsars and MSPs. According to the ATNF database v.~1.59\footnote{http://www.atnf.csiro.au/research/pulsar/psrcat}~\citep{atnf} the longest period binary discovered by the timing techniques has a period of 20-30 years (J2032+4127 \citealt{2015MNRAS.451..581L}), although some pulsars are known to have wider companions detected through a combination of different techniques. This is the case of PSR J1024-0719 which has an orbital period of more than 200~yrs according to \cite{2016MNRAS.460.2207B} and is in range 2-20~kyr according to \cite{2016ApJ...826...86K}.

%So far, a pulsar J2032+4127 with a longest orbital period discovered by means of the timing techniques has period of 20-30 years \citep{2015MNRAS.451..581L}.
Close triples, such as the PSR J0337+1715 discovered by \cite{2014Natur.505..520R} with an orbital period of third companion of $\approx 327$~days do not probe the lowest natal kicks because the third component of the PSR J0337+1715 could have been left bound to the system even after a natal kick with magnitude of 400~km~s$^{-1}$ \citep{2014ApJ...781L..13T}.
The LMXB V* V1727 Cyg (known as 4U 2129+47) seems to have a third component with an orbital period of $> 175$~d \citep{1989ApJ...341L..75G, 2008A&A...485..773B,2009ApJ...706.1069L}. In this case too small natal kicks are not restricted (see \cite{2011ApJ...734...55P} for a discussion of possible channels for the formation of MSP J1903+0327 through triple stellar evolution).
%If some NS were formed in wide binaries and received no natal kick and lose less than a half of the total binary mass in the episode of the supernova explosion, the wide binary could survive.

It might happen in rare cases that the magnitude and orientation of a strong natal kick is such that it allows for the formation of extremely eccentric wide binary with a large semi-major axis. However, our binary population synthesis models (see Appendix~\ref{s:population}) show that the probability for a chance formation of wide-binaries post-natal-kick is very low, $\approx 10^{-4}$~cases. Our small sample size of $\approx 50$ NSs, therefore makes it unlikely to find kick-formed wide-binaries.

We propose to use ultra-wide binaries to probe the weakest natal kicks (a few km~s$^{-1}$ or even a fraction of km~s$^{-1}$) by searching for wide astrometric pairs with common parallax and proper motion where one of the components is XRB, MSP or normal radio pulsar. This study follows to some extent the idea by \cite{2018MNRAS.480.4884E} to probe the possible natal kick velocity of white-dwarfs. In order to identify wide pairs with NS/BH compact objects components we use the Gaia second data release following the same principles we use for the study of wide companions to massive OB stars. These methods are described in the following. 
%We use the method by \cite{1988Ap&SS.142...17L} mostly following the prescription provided in \cite{2007AJ....133..889L}. 

\section{Methods}
\label{s:method}
For each of our objects (O/B star, HMXB, LMXB, pulsar or MSP) we select from the Gaia database all the stars found at angular distance up to 
\begin{equation}
\theta_\mathrm{max} \; [\mathrm{arcsec}] = 2\times 10^2 \; \varpi \; [\mathrm{mas}]     
\end{equation}
from our primary target. This choice is motivated by a fact that the number of chance aligned stars grows fast at distances  $>2\times 10^5$~a.u. 
We keep only stars with measured parallax and proper motion with relative errors which are less than a third of their value. We also check the quality of the five-parameters astrometric solutions for which we used a criterion similar to \cite{2018A&A...616A...2L}:
\begin{equation}
\sqrt{\frac{\chi^2}{\nu'-5}} < 1.2\; \mathrm{max}\left(1, \exp(-0.2[G-19.5])\right)  
\end{equation}
where $\chi^2$ is \texttt{astrometric\_chi2\_al} and $\nu'$ is \texttt{astrometric\_n\_good\_obs\_al} from the Gaia database.

For each pair we compute following values: angular separation at the sky $\Delta \Theta$, difference in parallax $\Delta \varpi$ and its error $\sigma_{\Delta \varpi}$, difference in proper motion $\Delta \mu$ and its error $\sigma_{\Delta \mu}$ and the possible expected difference in proper motion due to orbital motion $\mu_\mathrm{orb}$. In these calculations we follow the technique described in \cite{2018MNRAS.480.4884E}. The angular distance at the sky is: 
\begin{equation}
\Delta \theta = \sqrt{(\alpha_\mathrm{p} - \alpha_*)^2\cos^2 \delta_\mathrm{p} + (\delta_\mathrm{p} - \delta_*)^2}
\end{equation}
where $\alpha_\mathrm{p}$, $\alpha_*$, $\delta_\mathrm{p}$ and $\delta_*$ are the right ascension and declination of the primary and its possible second/third component, respectively. 
The difference in parallax is:
\begin{equation}
\Delta \varpi = |\varpi_\mathrm{p} - \varpi_*|    
\end{equation}
where $\varpi_\mathrm{p}$ and $\varpi_*$ are the parallaxes of the primary and the distant candidate companion, respectively.
The error of the parallax difference is:
\begin{equation}
\sigma_{\Delta \varpi} = \sqrt{\sigma_{\varpi, p}^2 + \sigma_{\varpi, *}^2 }
\end{equation}
The difference in proper motions is:
\begin{equation}
\Delta \mu = \sqrt{(\mu_{\alpha\mathrm{p}} -  \mu_{\alpha *})^2 + (\mu_{\delta\mathrm{p}} -  \mu_{\delta *})^2}    
\end{equation}
where the proper motion in the right ascension direction is $\mu_\alpha$ and in declination is $\mu_\delta$.
The error in the proper motion difference is:
\begin{equation}
\sigma_{\Delta \mu} = \frac{1}{\Delta \mu} \sqrt{(\sigma_{\mu \alpha p}^2 + \sigma_{\mu \alpha *}^2)\Delta \mu_\alpha^2 + (\sigma_{\mu \delta p}^2 + \sigma_{\mu \delta *}^2)\Delta \mu_\delta^2}    
\end{equation}
where $\sigma_{\mu \alpha p}$ and $\sigma_{\mu \alpha *}$ are errors of the proper motion in the right ascension direction for the primary and secondary stars.
The possible expected difference in proper motion due to orbital motion is computed as:
\begin{equation}
\mu_\mathrm{orb} \approx 0.53~\mathrm{mas}~\mathrm{year}^{-1} \left(\frac{\varpi_p}{\mathrm{mas}}\right)^{3/2} \left(\frac{\Delta \theta}{\mathrm{arcsec}}\right)^{-1/2}  
\end{equation}
The value is computed as maximum proper motion due to motion on circular orbit around a star with a mass $7M_\odot$:

We use the following conditions to identify the binaries:
\begin{enumerate}
    \item $\Delta \varpi < 2 \sigma_{\Delta\varpi}$
    \item $\Delta \mu < (\mu_\mathrm{orb} + 2 \sigma_{\Delta \mu})$
    \item $\sigma_{\Delta\varpi} < 0.12$~mas
    \item $\Delta \mu < 3  \mu_\mathrm{orb}$
    \item $\sigma_{\Delta\mu} < 0.16$~mas~year$^{-1}$
\end{enumerate}

These conditions are implemented as an ADQL request, see Appendix \ref{a:adql}. The first two conditions allow us to select stars which are close in phase-space. The last three conditions are set to remove a vast number of chance alignment of unassociated stars. These conditions were chosen using a displaced sample and visual inspection, as we describe in the following Section. The list of binaries with massive primary star identified by this method contains a significant number of members of open clusters. We describe the identification procedure for open clusters in the Section~\ref{s:clusters}. 

%For a large angular separation,  $\Delta \theta = 50''$ and a distance of 2~kpc the expected difference in proper motions (for stars to be physically bound) is $\delta \mu=0.03$~mas/year, comparable to the uncertainty of $\sigma_\mu = 0.04$~mas/year for relatively bright stars. This proper motion difference corresponds to $\approx 0.3$~km/s. On the other hand, the mean error in the proper motion measurement can be  quite large $\sigma \mu \approx 0.23$~mas/year. Therefore, this method (given the current second Gaia data release) is most sensitive to angular separations of $\Delta \theta < 25''$ corresponding to $5\times 10^4$~a.u. at 2~kpc. The smallest angular separation which can be found in the current Gaia data set is of the order of an arcsecond corresponding to an orbital separation of $2\times 10^{3}$~a.u. at a distance of 2~kpc.

%The use of $\Delta \theta$ and $\Delta \mu$ only  does not set any restrictions on the parallax difference, therefore any identified pair can be checked based on the parallax. We assume that the parallax difference between the main star and the secondary star should be less than $\Delta \varpi <1-3$~$\sigma_\varpi$, i.e. a typical  parallax difference in the range 0.04-0.12~mas. Another restriction is the transverse velocity difference between the main star and its possible distant companion, we assume that it should be less than $\Delta v < 0.5$~km~s$^{-1}$.

 \subsection{Chance alignment considerations}
The distribution of projected separations and the inferred binary fraction are affected by chance alignments of background stars. The main challenge in using this method is therefore to identify a boundary between a chance alignment identification and the identification of actual physical pairs.

First we visually inspect the sample using only $\Delta \theta$, $\Delta \mu$ and $\Delta \varpi$ following \cite{2007AJ....133..889L} to check if any binaries can be found in the sample. The distribution of these values for the O/B sample is shown in Figure~\ref{f:ver_separation} (right panel). We then change the declination of each O/B star in our sample by adding $50'$ (taking into account large parallax of $\varpi=16$~mas for some close massive stars) and plot the angular separations and the difference in the proper motion again, see Figure~\ref{f:ver_separation} (left panel). A clear presence of stars in the left lower corner of the plot convinces us that the sample contains real binaries. At this point we choose a values of 0.12~mas as the critical value for the error in parallax difference ($\sigma_{\Delta\varpi}$), 0.16~mas~year$^{-1}$ as the critical value for the error in the proper motion difference ($\sigma_{\Delta\mu}$) and the requirement that $\Delta \mu<3\mu_\mathrm{orb}$. The exact choice is somewhat arbitrary; and we decided on these values as to identify the binaries seen by eye in Figure~\ref{f:ver_separation}.

In order to estimate the chance alignment we run the complete search for binaries with chosen parameters for all massive stars, by first shifting their declination to $50'$ in positive direction and then shifiting to a negative direction. We identified 0 chance aligned pair in both cases.

\subsection{Cluster identification}
\label{s:clusters}
Many massive stars belong to open clusters or form OB associations. In such associations each star could potentially have many identified false wide companions. To simplify the analysis we therefore consider only field stars and discard stars in stellar clusters/associations in our anslysis.

In order to identify the field vs. cluster stars we make use of the open cluster catalogue by \cite{2018A&A...618A..93C} who identified 1229 open clusters using the second Gaia data release. For each O/B star with identified wide pairs we find all the clusters in the same region of the sky up to 1 degree around this star. An  O/B star is considered by us to be a member of a cluster if its parallax is within 0.3~mas from the cluster parallax and its proper motion is within 2~mas~year$^{-1}$ from the cluster mean proper motion. Any star with an identified potential host cluster is then discarded from our analysis of wide binaries.
%We also searched for common proper motion and parallax pairs using radius of 5~pc and the method described in the previous section.
%The results of our wide binary search together with cluster identification is shown in Table~\ref{t:b_candidate_cluster}. 

In general this procedure removes all the stars which were otherwise found to have large number of detected companions, with only a single exception of the HD 183058 stars which is found to have  17 companions (separations less than 5~pc) around it with similar parallaxes and proper motions. We additionally checked the \texttt{WEBDA} catalogue \citep{1995ASSL..203..127M} for this star. The star lacks known cluster association. It therefore suggests that our analysis identifies a new previously unknown open cluster or a stellar stream. We exclude this star and its companions from our following analysis.

\section{Ultra-wide O/B-stars}
\label{s:verif}
  It is known that up to a half of all massive stars are born in binaries \citep{sana2012} of any separation (at least in the LMC massive clusters). Around $\sim 10$ per cent of low-mass stars are in triples and this fraction reaches even up to $35$~per cent for B stars \citep{moe2017}. The majority of stable triple stars form hierarchical systems with an inner compact pair and a distant companion. It is known that some triples survive the X-ray binary stage (see examples in Section~\ref{s:ns}). However,  previous studies typically focused on binaries with smaller separations than considered here, and our study therefore complements our current understanding of the multiplicity of massive stars.

\begin{figure*}
\begin{minipage}{0.48\linewidth}
	\includegraphics[width=\columnwidth]{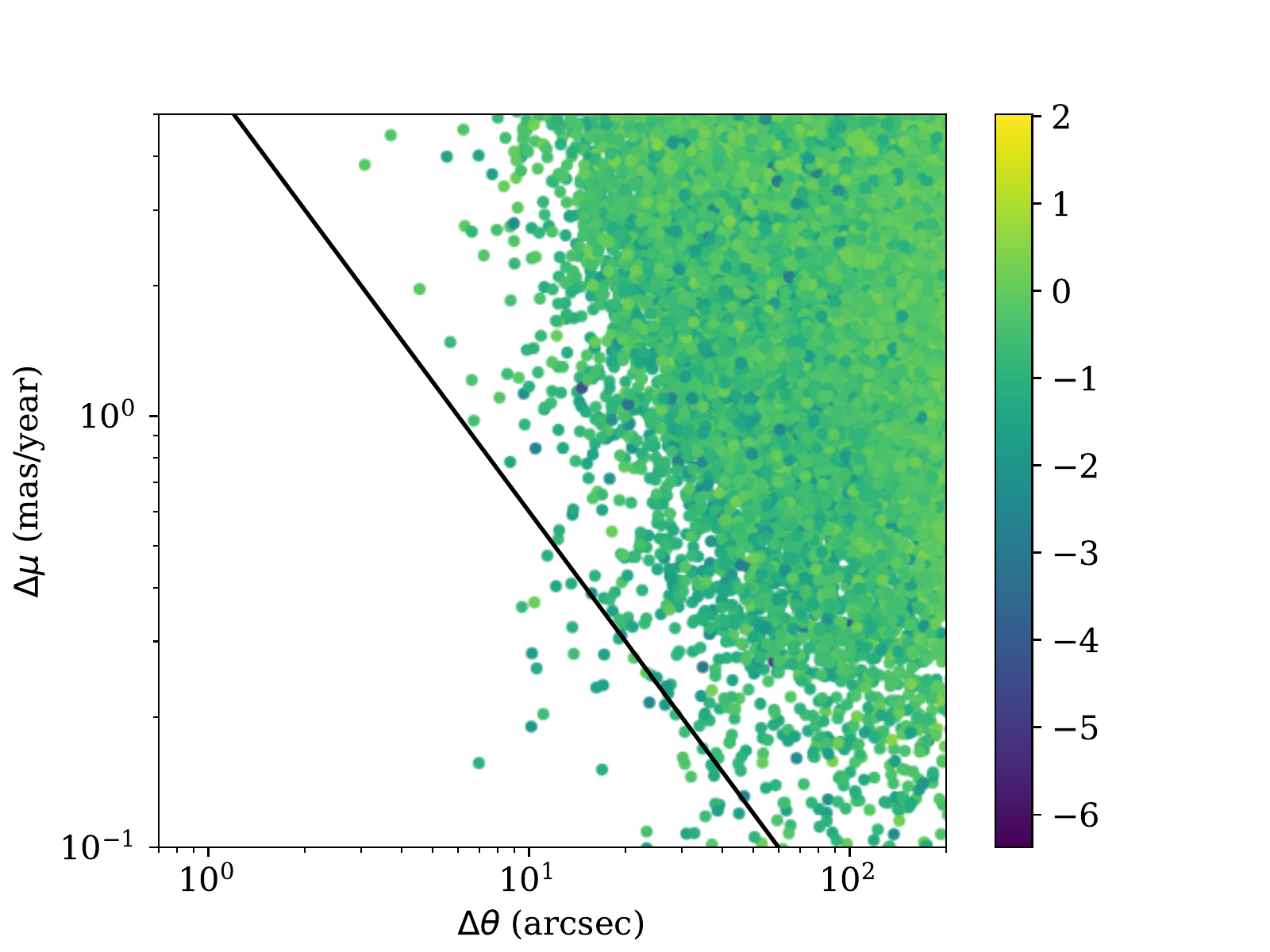}
\end{minipage}
\begin{minipage}{0.48\linewidth}
	\includegraphics[width=\columnwidth]{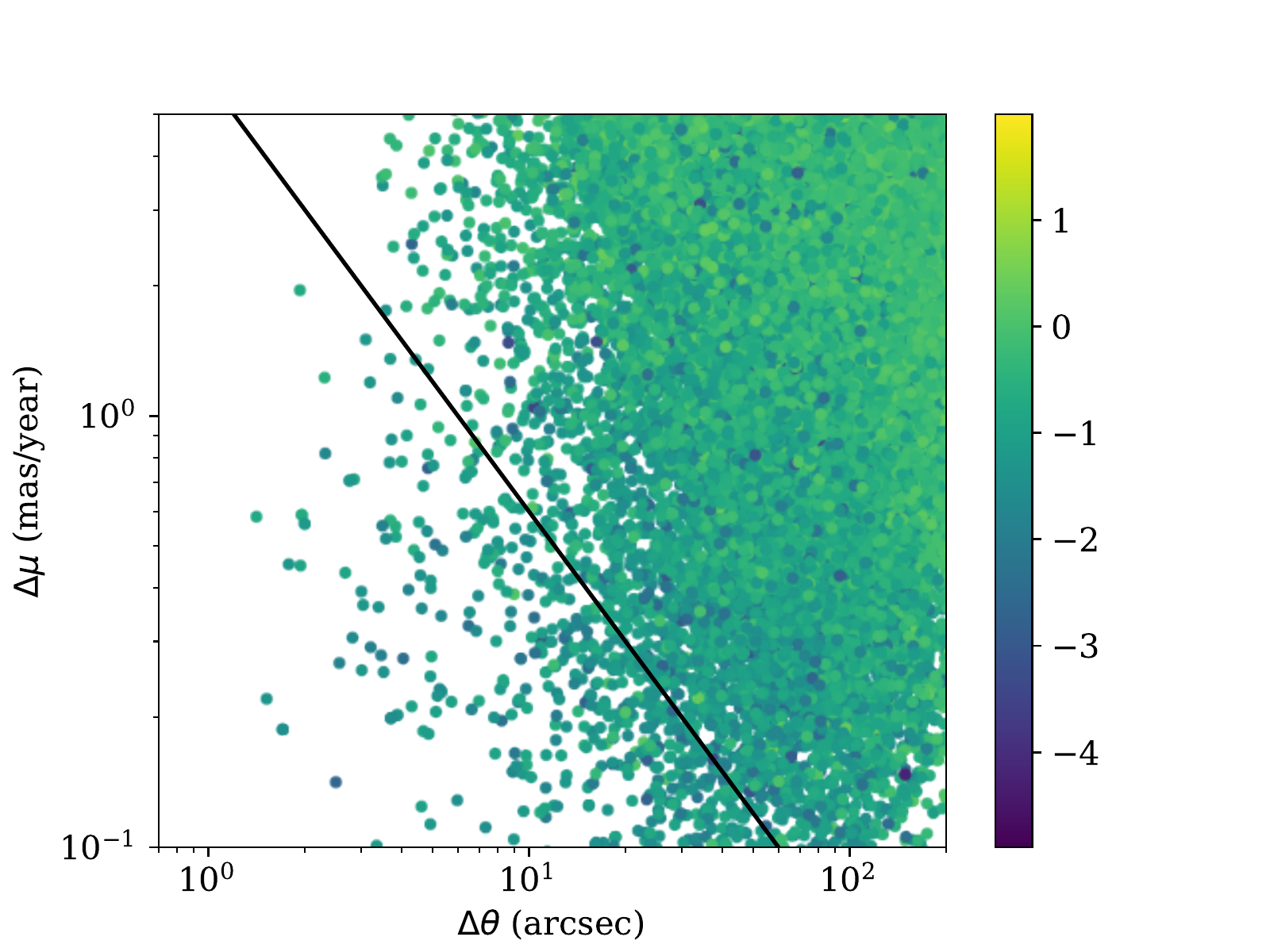}
\end{minipage}
\caption{The difference in proper motion vs. the angular separation for the O/B-stars sample. Left panel: All the objects are shifted by $50'$ in the direction of declination as to exclude  any possible real wide binaries from the mock-sample. Right panel: The observed distributions.  The color shows the logarithm of the parallax difference. Black solid line assists comparison between the misplaced and observed sample. }
    \label{f:ver_separation}
\end{figure*}

\begin{figure*}
\begin{minipage}{0.48\linewidth}
	\includegraphics[width=\columnwidth]{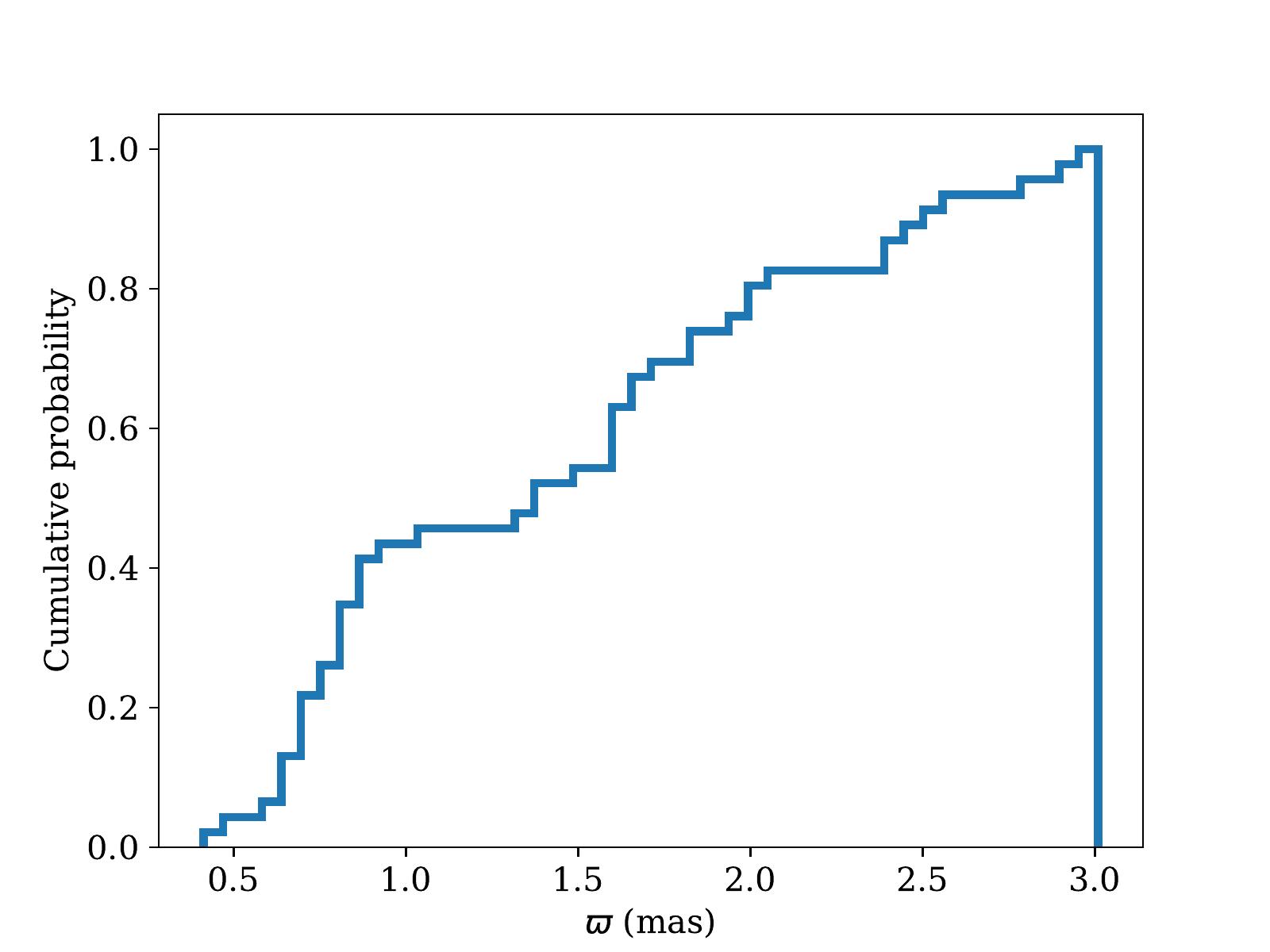}
\end{minipage}
\begin{minipage}{0.48\linewidth}
	\includegraphics[width=\columnwidth]{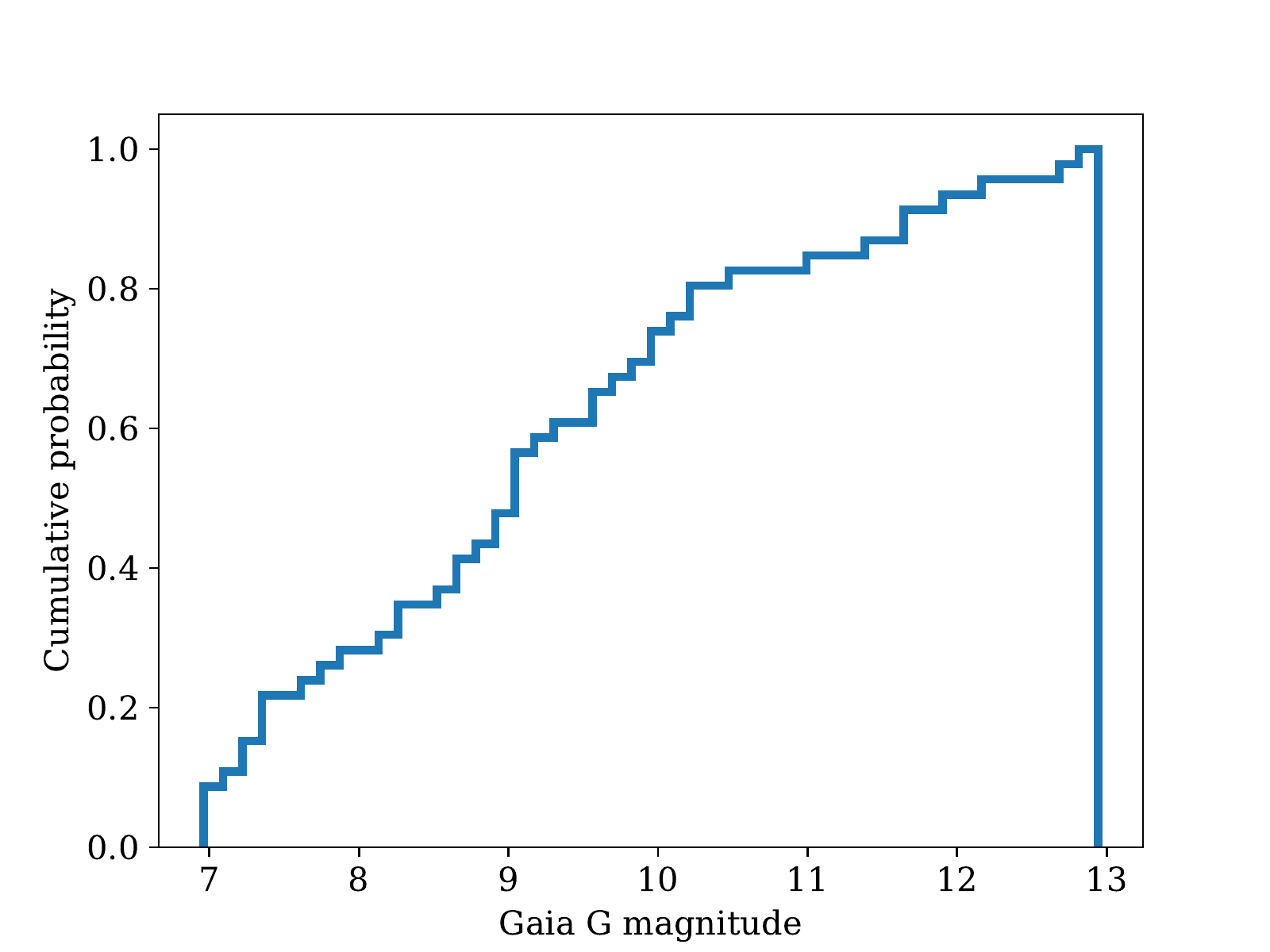}
\end{minipage}
\caption{Cumulative distribution of parallaxes (left panel) and apparent G magnitude (right panel) for primary O/B stars with wide companions. }
    \label{f:OB_host}
\end{figure*}

For our analysis we use the second Gaia data release \citep{gaia_mission, gaia_dr2} as well as the Simbad database and catalogue of Galactic O stars \citep{goss1,goss2,goss3}. Only O/B stars (O-B5 spectral types, see Appendix~\ref{a:adql} for the ADQL requests) with measured V magnitude are chosen.
Our initial list contains 1674 (B0-B5 stars) and 588 (O stars). After removing stars with problematic astrometric measurements according to the Gaia database we keep 1937 stars. 
%The list contains only seven stars with O spectral type, most stars are B0-B5.
Applying the method described above we get the results in Figure~\ref{f:ver_separation}. In our search we identified 69 reliable wide binaries, see Table~\ref{t:b_candidate}. We plot the distribution of parallaxes and apparent Gaia G magnitudes in Figure~\ref{f:OB_host}.

\begin{figure*}
\begin{minipage}{0.48\linewidth}
	\includegraphics[width=\columnwidth]{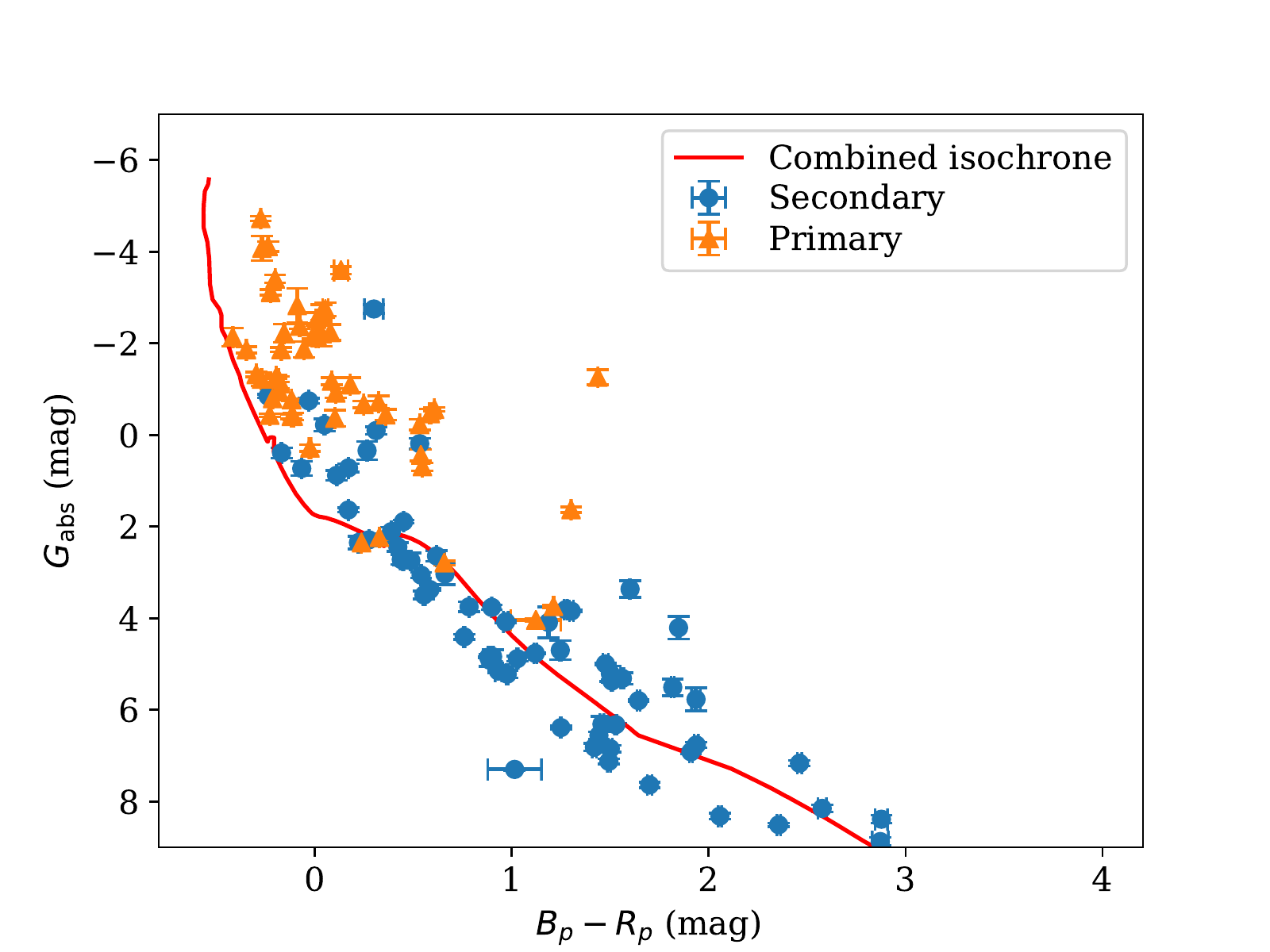}
\end{minipage}
\begin{minipage}{0.48\linewidth}
	\includegraphics[width=\columnwidth]{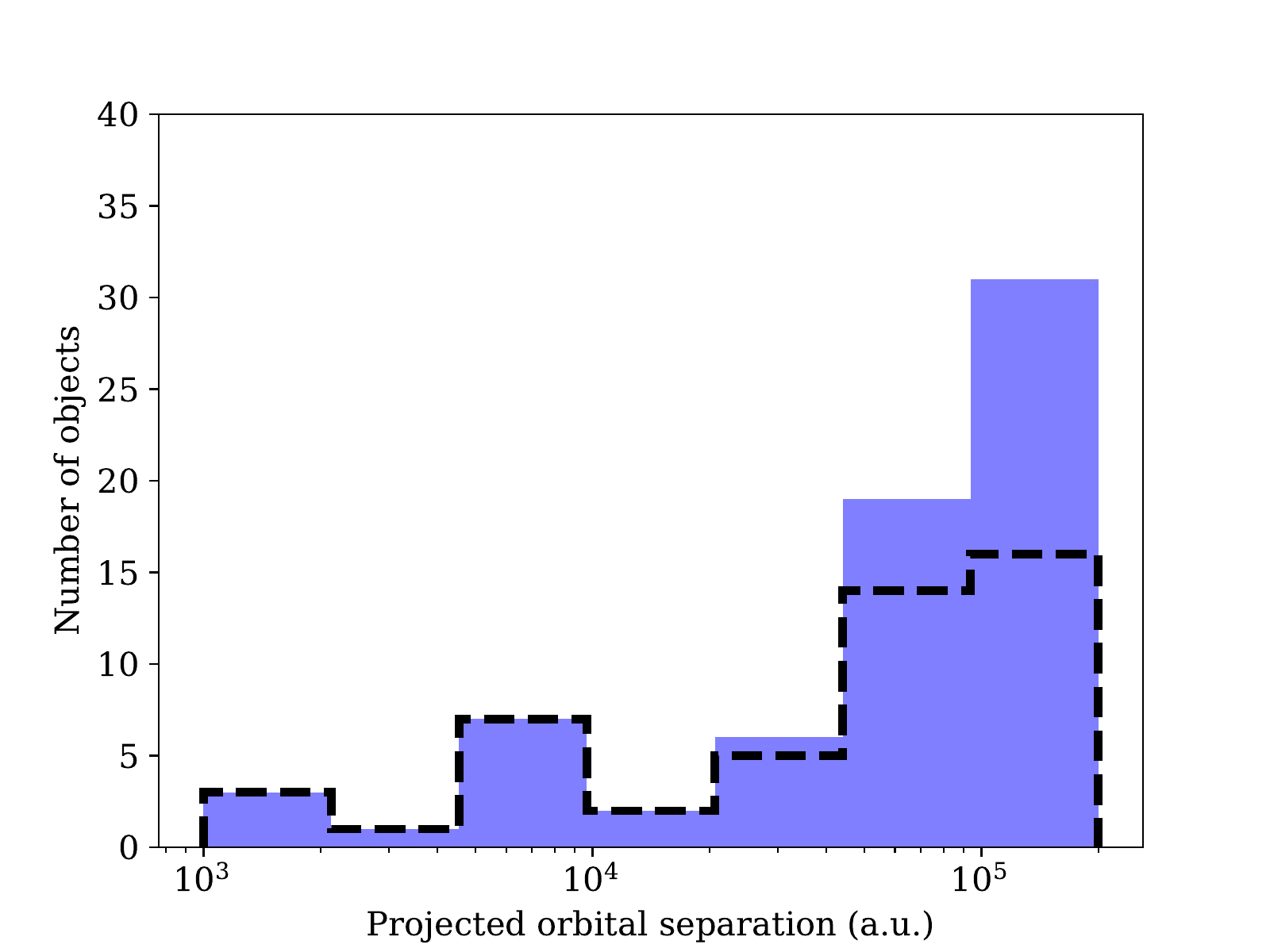}
\end{minipage}
\caption{The Hertzsprung - Russell diagram (left panel) and the distribution of the observed spatial separations (right panel) for the wide astrometric pairs found in the in the O/B-stars sample. 
The blue (top filled) histogram corresponds to the separation of all candidates from the Table~\ref{t:b_candidate}, and the black dashed  histogram corresponds to the cases where  only the closest component was selected in each case where more than a single companion was identified. 
The isochone is combined from 2~Myr and 10~Myr isochones for stars more massive and less massive than $3.5~M_\odot$ respectively.
%The dot-and-dashed red line show the false-positive rate found in 100 Monte Carlo simulations and the black dotted line shows false positive rate found in the $10'$ shifted sample.
%Vertical solid line shows limit $3.3\times 10^5$~a.u., the disruption timescale by molecular clouds is shorter than 10~Myr for orbital separations larger than this value. 
}
    \label{f:ver_hist}
\end{figure*}

\begin{figure}
%\begin{minipage}{0.48\linewidth}
	\includegraphics[width=\columnwidth]{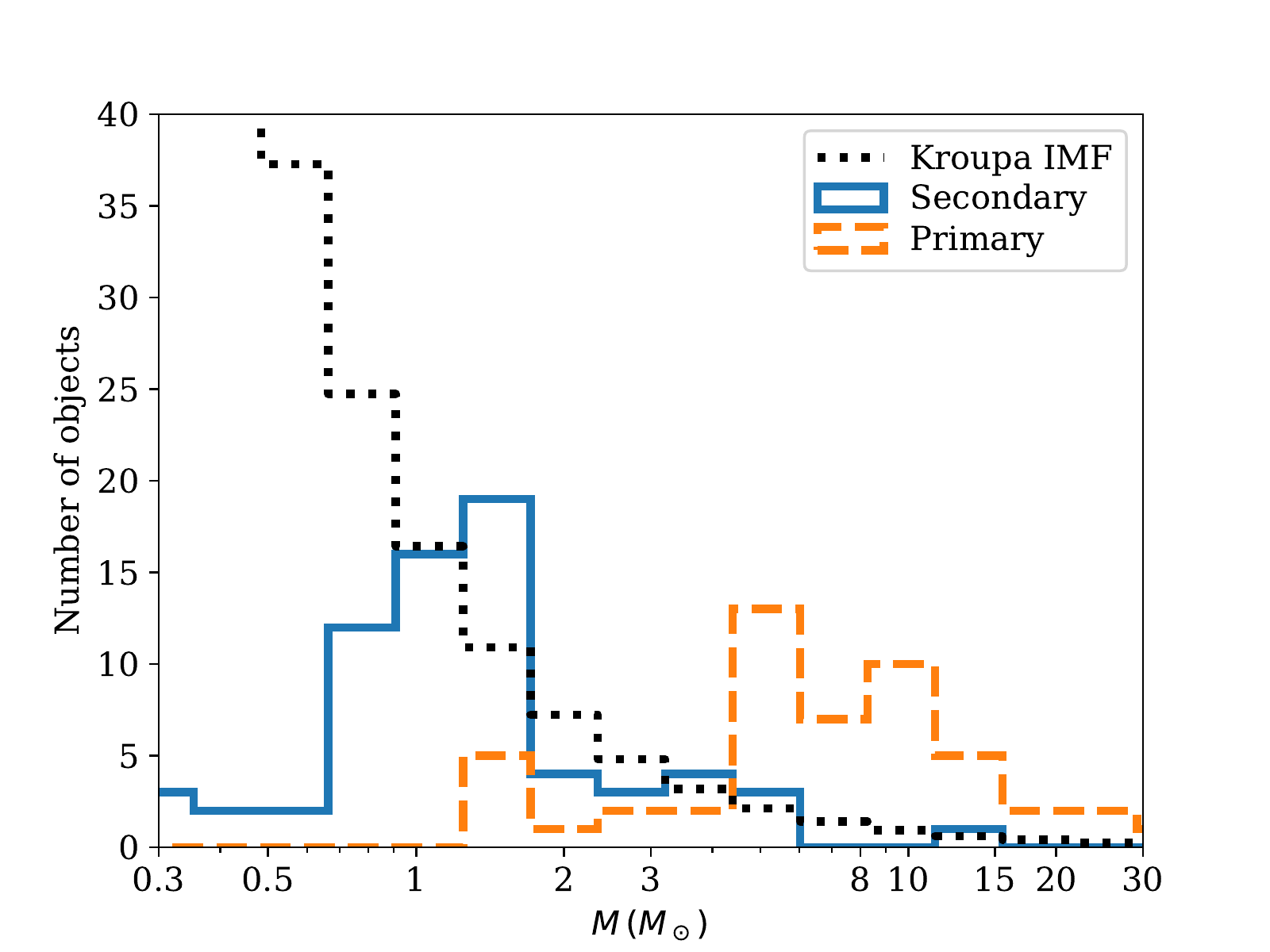}
%\end{minipage}
\caption{Distribution of masses of primaries and secondaries found in the O/B stellar sample. The dotted black line shows the Kroupa IMF consistent with mass distribution of secondaries more massive than 1.3~$M_\odot$. }
    \label{f:masses}
\end{figure}

%In our simulations, see Fugire~\ref{f:ver_separation}, in a shifted sample we found 18 good candidates. Therefore up to $38$~per cent of the sample presented in Table~\ref{t:b_candidate} could appear due to a chance alignment. Careful analysis of radial velocities which will be available in the next Gaia data release will make it possible to prepare wide binary sample better. 

%All the wide pairs we find have parallax differences below $0.09$~mas (except V* V1061 Tau and HD 158724) while the typical parallax is $\sim 0.5$~mas. The velocity difference is below $0.5$~km/s and the projected orbital separations are in the range  $\approx 10^3-$few$\times10^5$~a.u. 
%Some objects have multiple possible wide pairs which might indicate that these stars belong to an OB association or that they are higher multiplicity systems. 
%Remarkably, we see 21 pairs with angular separations $2.5\leq \Delta \theta < 25$~arcsec. Such pairs are not found at all in the case of HMXBs or normal pulsars. 

The distribution of projected separation is presented in Figure~\ref{f:ver_hist}. We also tried to exclude all distant pairs to stars which already have a component at smaller angular separation. 
%In this case we get a distribution which seems to be similar with flat lognormal. 
%An excess of components at separations of few$\times 10^5$~a.u. is probably caused by lack of radial velocity measurements. Future observations could confirm the nature of these wide systems. 
%The wide binarity rate is then $48/449\approx 0.11$ if we do not exclude multiple associations and $24/449\approx 0.05$ if we exclude all multiple associations. It is worth to notice that some of multiple associations are real because stars could form high multiplicity systems. 

When we run the cluster identification for the whole list of 1937 O/B stars we find that only 353 of them are in clusters. This seemingly small fraction of $\approx 18$~per~cent is mostly due to the short lifetimes of most stellar clusters and associations, as we briefly discuss in the following.

Most stars (70-90 percents) are initially formed in stellar cluster/associations embedded in giant molecular clouds \citep{2003ARA&A..41...57L}. Such clusters  are not seen in optics and undetectable by Gaia. Only 4-7 per cent of embedded clusters survive the gas expulsion phase to become visible as open clusters or OB associations. In particular, the lifetime of small clusters (up to 1000 stars) is about 10-20~Myr, with the first 10~Myr spent in the gas-embedded stage. Moreover, since our sample is dominated by B stars (which lifetime is longer than the typical lifetime of visible clusters) it is not surprising to find the majority of observed OB stars in our sample in the field.

Our technique might also be a subject to following caveats: (1) some of the stars are in clusters/associations which have not been previously identified (see previous section for example of HD 183058) and (2) although some of the OB stars are not hosted by currently identified clusters, they may still have gravitationally unbound neighbors in phase space from a birth cluster in the process of dissolving.

The inferred multiplicity rate for the field O/B stars is $69 / (1937-353) \approx 4.4$~per cent.
We estimate the variance for the mean estimator as $\sqrt{\hat p (1-\hat p) / n} \approx 0.5$ per cent, we therefore see a  multiplicity rate of $0.044\pm 0.005$. We also estimate this fraction independently for O and B stars. For O stars outside of open cluster ($48.5$ per cent of all O stars in our sample) the multiplicity fraction is $6/198 = 3\pm 1$ per cent. For B stars outside of open cluster (90 per cent of all B stars in our sample) the multiplicity fraction is 4.5 per cent.  
%The  $0.5$\% uncertainty is also compatible with the difference which we get between the Monte Carlo (synthetic) and the shifted (observed) sample estimates for false positive probability since the difference is $(18-14)/449\approx 1$\%.  

 In our sample, due to the luminosity limit, we are only sensitive to wide components of O, B, A, F, G and partly K stars (see the Hertzsprung-Russell diagram in Figure~\ref{f:masses}). In order to plot the Hertzsprung-Russell diagram we correct the magnitudes for extinction using the \cite{2018MNRAS.478..651G} map. In order to convert from $E(B-V)$ to $A_G$ and $E(Bp-Rp)$ we use $R=3.1$, $A_g/A_v = 0.9$ and $E(Bp - Rp) / E(B-V) = 1.5$ and use a general correction of 0.884 for all the reddening values obtained from the map. If the location of the star is not covered by the map, we decide not to correct for the reddening. We infer the masses using combined Padova stellar isochrones (stellar tracks from \citealt{2017ApJ...835...77M}) for a stellar age of 2 and 10~Myr for stars more and less massive than $3.5~M_\odot$ with the Gaia DR2 photometric system as described by \cite{2018A&A...616A...4E}.

 The inferred mass distribution of primary and secondary stars is plotted in Figure~\ref{f:masses}. Some primaries might be misclassified, for example CD-49  4263 seems to have mass $1.4~M_\odot$ which is below B9. There are 6 primary stars with inferred masses less than 2~$M_\odot$ in our sample. These are CD-49  4263, 2MASS J05065551-0321132, HD 225190, HD 344313, TYC 3986-3080-1 and V* V1671 Cyg.

 Another effect which we see in the mass distribution is that the low-mass stars are missing ($M < 0.5M_\odot$) as expected from our selection bias for detecting only higher luminosity companions. In order to get a rough estimate for the total multiplicity fraction, one can consider the non-trivial assumption that the companion mass-function continues to follow the Kroupa initial mass function (IMF) \citep{2001MNRAS.322..231K} down to low-mass stars, below our detection limit. Under this assumption we identify our lowest detectable companion mass $M_0$, above which the survey is complete. We then normalize the Kroupa IMF as to get the exact number of stars with masses more than $M_0$. The number of stars with masses less than $M_0$ can then be extrapolated from the assumed companion IMF. Since the exact completeness cut-off is difficult to asses, we consider $M_0$ in the range of 0.8-1.5~$M_\odot$ and find the corresponding multiplicity fraction ranges between 22 to 32 per cent. Roughly we can say that the extrapolated multiplicity fraction is about $27\pm 5$ per cent. In reality the systematic uncertainty arising from our assumption of the companion IMF to continue to low-mass values, could be much larger.

 %If the companion mass function follows the Kroupa mass function we would expect to have $\approx 6$ times more stars in the mass range $0.01-0.5~M_\odot$ than in the observable mass range of $0.5-16~M_\odot$. If we follow this assumption we can estimate the overall wide multiplicity fraction for OB stars with stellar companions of any mass to be up to $\sim48\pm6$ per cent. 
For comparison, the wide multiplicity rate for low mass primaries is $9.1\%\pm 1.6\%$ found for multiplicity of primary G dwarfs with separations larger than $10^3$~a.u. by \cite{2007AJ....133..889L} and is dominated by M-dwarf companions. In other words the ultra-wide binary fraction of massive stars might be 2-3 times higher than that of low-mass stars, a result consistent with theoretical expectations for dynamically-formed wide binaries where the ultra-wide binary fraction is expected to be much higher than that of low-mass stars \citep{Per+12}.  Since previous estimates of the multiplicity fraction of massive stars were limited to smaller separations than those considered here, this result hints a significant increase in the inferred multiplicity fraction of massive stars, with the majority being part of triple or higher multiplicity systems.

\begin{table*}
    \centering
    \begin{tabular}{lrrrrrrcrccc}
        \hline
        Name O/B star &  Gaia primary & Gaia secondary & $\Delta \theta$  & $\Delta\varpi \pm \sigma_{\Delta\varpi}$ & $\Delta \mu\pm \sigma_{\Delta\mu}$ & $A$ \\
                    & Gaia DR2 name & Gaia DR2 name  & (arcsec)           & (mas)                                          & (mas year$^{-1}$)                        & (a.u.)\\
        \hline
2MASS J05065551-0321132      &  3213014868922447232  &  3213015148092614656  &  120.9  & $ 0.116  \pm  0.071 $ & $ 0.195  \pm  0.096  $ &  48071.9  \\
2MASS J05065551-0321132      &  3213014868922447232  &  3213016007087564160  &  229.9  & $ 0.088  \pm  0.046 $ & $ 0.316  \pm  0.085  $ &  91444.7  \\
2MASS J05065551-0321132      &  3213014868922447232  &  3213010750047595520  &  281.1  & $ 0.022  \pm  0.042 $ & $ 0.065  \pm  0.076  $ &  111789.6  \\
2MASS J05065551-0321132      &  3213014868922447232  &  3213010509529424384  &  389.5  & $ 0.058  \pm  0.074 $ & $ 0.05  \pm  0.094  $ &  154913.3  \\
2MASS J05065551-0321132      &  3213014868922447232  &  3212957664251821824  &  443.3  & $ 0.066  \pm  0.044 $ & $ 0.06  \pm  0.082  $ &  176322.3  \\
BD+50   377                  &  358015279655161088  &  358009442797341568  &  119.4  & $ 0.101  \pm  0.084 $ & $ 0.139  \pm  0.135  $ &  48529.1  \\
BD+50   377                  &  358015279655161088  &  358009339718126720  &  136.4  & $ 0.085  \pm  0.063 $ & $ 0.24  \pm  0.107  $ &  55417.4  \\
CD-44 11306                  &  5964099137533440640  &  5964099343691734016  &  46.8  & $ 0.091  \pm  0.067 $ & $ 0.117  \pm  0.102  $ &  60453.3  \\
CD-44 11306                  &  5964099137533440640  &  5964098897015324928  &  103.2  & $ 0.091  \pm  0.063 $ & $ 0.085  \pm  0.12  $ &  133261.6  \\
CD-44 11306                  &  5964099137533440640  &  5964099446770967424  &  153.1  & $ 0.117  \pm  0.066 $ & $ 0.058  \pm  0.119  $ &  197566.0  \\
CD-49  4263  &  5325671971449181696  &  5325671971449183488  &  16.5  & $ 0.013  \pm  0.055 $ & $ 0.047  \pm  0.099  $ &  39505.6  \\
CD-49  4263  &  5325671971449181696  &  5325672761723170176  &  52.6  & $ 0.04  \pm  0.045 $ & $ 0.027  \pm  0.079  $ &  126113.8  \\
HD  20508                    &  462485965385240832  &  462482838649050496  &  124.4  & $ 0.08  \pm  0.056 $ & $ 0.12  \pm  0.076  $ &  151438.2  \\
HD  21650                    &  237762796737339776  &  237739191597107456  &  139.5  & $ 0.038  \pm  0.109 $ & $ 0.203  \pm  0.159  $ &  46341.3  \\
HD  24190                    &  218790861034319744  &  218791376433618304  &  156.0  & $ 0.115  \pm  0.058 $ & $ 0.414  \pm  0.123  $ &  64369.3  \\
HD  33090                    &  3414941966067063552  &  3414941966069295744  &  9.0  & $ 0.059  \pm  0.065 $ & $ 0.104  \pm  0.105  $ &  4417.8  \\
HD  45677                    &  2999967452909563136  &  2999967074952442624  &  29.4  & $ 0.01  \pm  0.083 $ & $ 0.089  \pm  0.135  $ &  18369.1  \\
HD  52533  &  3108642147107724544  &  3108641773451669376  &  11.3  & $ 0.01  \pm  0.071 $ & $ 0.117  \pm  0.107  $ &  21915.9  \\
HD 154538                    &  5926578921730646144  &  5926578986111699712  &  9.9  & $ 0.034  \pm  0.071 $ & $ 0.469  \pm  0.111  $ &  6942.3  \\
HD 155606                    &  5953022382536807424  &  5953021690999065856  &  125.5  & $ 0.118  \pm  0.084 $ & $ 0.083  \pm  0.131  $ &  137246.6  \\
HD 157020                    &  5953355220976166016  &  5953355225327896064  &  43.1  & $ 0.039  \pm  0.085 $ & $ 0.111  \pm  0.14  $ &  52858.9  \\
HD 158724                    &  5974604764982749440  &  5974604558791120640  &  8.9  & $ 0.11  \pm  0.074 $ & $ 0.203  \pm  0.1  $ &  6418.4  \\
HD 158724                    &  5974604764982749440  &  5974604760652216320  &  44.0  & $ 0.04  \pm  0.087 $ & $ 0.219  \pm  0.146  $ &  31910.9  \\
HD 162418                    &  4036689403295912704  &  4036688548653697664  &  75.8  & $ 0.082  \pm  0.081 $ & $ 0.08  \pm  0.103  $ &  90670.4  \\
HD 162540                    &  4063721931768310528  &  4063721893097768320  &  56.8  & $ 0.102  \pm  0.079 $ & $ 0.259  \pm  0.147  $ &  32324.1  \\
HD 162540                    &  4063721931768310528  &  4063721313293010560  &  229.0  & $ 0.17  \pm  0.088 $ & $ 0.161  \pm  0.155  $ &  130264.2  \\
HD 163393                    &  4056195293667038592  &  4056195224947573760  &  75.4  & $ 0.106  \pm  0.103 $ & $ 0.025  \pm  0.152  $ &  129469.4  \\
HD 164950                    &  4064121913535676160  &  4064125143351139840  &  138.3  & $ 0.04  \pm  0.062 $ & $ 0.05  \pm  0.096  $ &  187113.1  \\
HD 167474                    &  4097483776326830720  &  4097483776326831744  &  23.7  & $ 0.03  \pm  0.076 $ & $ 0.106  \pm  0.115  $ &  33060.0  \\
HD 169675                    &  4096208514732308864  &  4096208372898138880  &  86.6  & $ 0.124  \pm  0.073 $ & $ 0.179  \pm  0.108  $ &  51630.3  \\
HD 170051                    &  4585303172501204480  &  4585315778226217728  &  393.7  & $ 0.032  \pm  0.05 $ & $ 0.247  \pm  0.068  $ &  135849.8  \\
HD 170051                    &  4585303172501204480  &  4584923566112188672  &  532.2  & $ 0.094  \pm  0.066 $ & $ 0.103  \pm  0.102  $ &  183648.8  \\
HD 170051                    &  4585303172501204480  &  4584939955707538560  &  541.2  & $ 0.063  \pm  0.049 $ & $ 0.141  \pm  0.082  $ &  186741.1  \\
HD 171348                    &  4079494769740363136  &  4079494769740368640  &  11.4  & $ 0.049  \pm  0.085 $ & $ 0.098  \pm  0.113  $ &  10912.7  \\
HD 177648                    &  4521438177054351104  &  4521414709351870720  &  12.4  & $ 0.033  \pm  0.058 $ & $ 0.396  \pm  0.08  $ &  7576.1  \\
HD 179126                    &  4264306623991981952  &  4264301577449116544  &  251.9  & $ 0.056  \pm  0.086 $ & $ 0.046  \pm  0.105  $ &  152349.1  \\
HD 185418                    &  4322021849109320320  &  4322021844762504064  &  12.5  & $ 0.058  \pm  0.048 $ & $ 0.22  \pm  0.068  $ &  9410.6  \\
HD 186363                    &  4239724640671288960  &  4239722922684335488  &  227.0  & $ 0.063  \pm  0.074 $ & $ 0.228  \pm  0.127  $ &  134856.5  \\
HD 213023A  &  2205471994328733568  &  2205471998624679040  &  1.8  & $ 0.053  \pm  0.058 $ & $ 0.452  \pm  0.074  $ &  1827.6  \\
HD 213023A  &  2205471994328733568  &  2205472307866244992  &  75.7  & $ 0.077  \pm  0.041 $ & $ 0.183  \pm  0.063  $ &  77397.1  \\
HD 217086  &  2207291656008154112  &  2207269562701436160  &  229.0  & $ 0.023  \pm  0.048 $ & $ 0.119  \pm  0.081  $ &  195042.4  \\
HD 217817                    &  2013630893446412672  &  2013618828875520512  &  161.8  & $ 0.035  \pm  0.081 $ & $ 0.195  \pm  0.13  $ &  79268.7  \\
HD 220598                    &  1913545407978693632  &  1913543483833350528  &  258.5  & $ 0.078  \pm  0.091 $ & $ 0.238  \pm  0.108  $ &  106662.6  \\
HD 225190                    &  420276126312352896  &  420276981005904128  &  161.6  & $ 0.112  \pm  0.11 $ & $ 0.364  \pm  0.14  $ &  63115.7  \\
HD 225190                    &  420276126312352896  &  420298253984619520  &  417.8  & $ 0.087  \pm  0.05 $ & $ 0.191  \pm  0.07  $ &  163181.3  \\
HD 225190                    &  420276126312352896  &  420298013466438912  &  495.7  & $ 0.025  \pm  0.118 $ & $ 0.209  \pm  0.151  $ &  193605.8  \\
HD 226609                    &  2072510634482203520  &  2072510634482203136  &  3.2  & $ 0.055  \pm  0.041 $ & $ 1.193  \pm  0.073  $ &  1149.9  \\
HD 235523                    &  2172465655999346432  &  2172465759071547648  &  91.4  & $ 0.037  \pm  0.045 $ & $ 0.251  \pm  0.074  $ &  49480.0  \\
HD 237121                    &  461841342334222592  &  461841101815943936  &  100.6  & $ 0.115  \pm  0.114 $ & $ 0.114  \pm  0.156  $ &  121025.4  \\
HD 248903                    &  3444121565857570560  &  3444122321771813632  &  68.8  & $ 0.053  \pm  0.087 $ & $ 0.036  \pm  0.126  $ &  166820.7  \\
HD 248903                    &  3444121565857570560  &  3444121290979664256  &  70.4  & $ 0.117  \pm  0.079 $ & $ 0.064  \pm  0.107  $ &  170759.8  \\
HD 248903                    &  3444121565857570560  &  3444122351833299968  &  81.1  & $ 0.113  \pm  0.082 $ & $ 0.028  \pm  0.108  $ &  196568.3  \\
HD 252532                    &  3425061463071277184  &  3425061458773888512  &  12.4  & $ 0.011  \pm  0.043 $ & $ 0.084  \pm  0.072  $ &  24446.6  \\
HD 282430                    &  158645612194039808  &  158645612195930624  &  4.2  & $ 0.018  \pm  0.081 $ & $ 0.395  \pm  0.148  $ &  6184.7  \\
HD 312640                    &  4095537262805235072  &  4095537017965216384  &  104.8  & $ 0.014  \pm  0.069 $ & $ 0.082  \pm  0.116  $ &  153882.7  \\
HD 314057                    &  4093933448976732160  &  4093933586410393216  &  117.0  & $ 0.087  \pm  0.101 $ & $ 0.1  \pm  0.138  $ &  129831.8  \\
HD 314057                    &  4093933448976732160  &  4093934381021176448  &  143.5  & $ 0.086  \pm  0.091 $ & $ 0.022  \pm  0.14  $ &  159232.5  \\
HD 328064                    &  5990524898748290048  &  5990545381493482880  &  129.2  & $ 0.105  \pm  0.071 $ & $ 0.024  \pm  0.091  $ &  189741.9  \\
HD 328907                    &  5964302267984622336  &  5967281158540457600  &  166.6  & $ 0.079  \pm  0.09 $ & $ 0.102  \pm  0.146  $ &  180995.2  \\
HD 329213                    &  5964098720895565440  &  5964099343691734016  &  160.4  & $ 0.077  \pm  0.071 $ & $ 0.069  \pm  0.131  $ &  170197.2  \\
HD 344313                    &  2019268531933154176  &  2019267093154924416  &  104.2  & $ 0.02  \pm  0.04 $ & $ 0.11  \pm  0.063  $ &  56755.5  \\
LB  3357                     &  4668594889152178304  &  4668594889152178560  &  6.6  & $ 0.019  \pm  0.028 $ & $ 0.208  \pm  0.054  $ &  9001.4  \\
TYC 3728-530-1               &  449218983208306304  &  449218983208307200  &  5.2  & $ 0.048  \pm  0.073 $ & $ 0.224  \pm  0.091  $ &  6515.5  \\
TYC 3986-3080-1              &  2005104898028283520  &  2005104936694779264  &  59.3  & $ 0.016  \pm  0.055 $ & $ 0.128  \pm  0.091  $ &  81001.3  \\
TYC 3986-3080-1              &  2005104898028283520  &  2005104799255829504  &  66.2  & $ 0.013  \pm  0.057 $ & $ 0.073  \pm  0.1  $ &  90435.5  \\
V* IS Nor                    &  5932588474413552512  &  5932588474413551488  &  2.5  & $ 0.002  \pm  0.046 $ & $ 0.141  \pm  0.07  $ &  1519.5  \\
V* V1671 Cyg                 &  2032810465079954944  &  2032810224561748480  &  160.5  & $ 0.103  \pm  0.068 $ & $ 0.239  \pm  0.117  $ &  104972.5  \\
V* V730 Cep                  &  2013474586693428096  &  2013474659716546688  &  150.1  & $ 0.004  \pm  0.088 $ & $ 0.106  \pm  0.128  $ &  72220.1  \\
V* V730 Cep                  &  2013474586693428096  &  2013474659716547968  &  154.0  & $ 0.037  \pm  0.065 $ & $ 0.183  \pm  0.1  $ &  74051.9  \\
    \hline
    \end{tabular}
    \caption{Wide binaries found in the O/B sample. $A$ is the projected orbital separation.}
    \label{t:b_candidate}
\end{table*}

\section{Ultra-wide NS/BH systems}
We identify HMXBs and LMXBs in the second Gaia data release based on their optical magnitude (the exact procedure is described in Section~\ref{s:xrb}). This approach does not work for normal and millisecond pulsars because of their faintness (the brightest object is the Crab pulsar with $V=16.6$~mag, see \citealt{2011AdSpR..47.1281M}). Instead of following the same procedure as for HMXBs and LMXBs, we try to identify normal pulsars and MSPs in the second Gaia data release based on position (circle of 3 arcsec around the radio position) and proper motion.  

We failed to find any reliable (relative errors of parallax and proper motion measurements less than 0.25) optical counterpart for isolated normal radio pulsars based on their coordinate and proper motion. 
We have identified five optical counterparts (angular separation less than $3''$, similar proper motions) to MSPs, see Table~\ref{t:ident_gaia}. These pulsars and their optical counterparts are the same as found in \cite{2018arXiv181206262M} and \cite{2018ApJ...864...26J}, except for J1732-5049 and J1910+1256. Pulsar J1732-5049 does not have the timing measurement for parallax \citep{rhc+16}, so it is difficult to verify this association. In the case of J1910+1256 the Gaia second data release provides only upper limit on the parallax and proper motion in the right ascension direction, so although we found the same association as in \cite{2018arXiv181206262M},  we excluded this object. In any case the number and quality of the Gaia measurements for MSPs counterparts prevent us from using these parallaxes and proper motions to search for wide pairs. 
%The only MSP reliably identified in the Gaia data set  is the J1023+0038 which is transitional pulsar which shows period of accretion as a member of LMXB \citep{2009Sci...324.1411A}.  
Therefore, instead, we use the measurements of the parallaxes and proper motions obtained from radio observations of pulsars to search for possible second/third components in the optical data from  Gaia. This approach is justified, since both the timing and the interferomertic proper motion measurements as well as the Gaia  measurements are all performed in the ICRF. 

%For normal pulsars this approach does not work since the optical magnitude of normal radio pulsars exceeds $V=20$~mag (with an . For majority of normal pulsars there is no optical detections. For MSPs we have identified only 16 objects based on similarity of parallax and proper motions. All but one system does not have good enough parallax and proper motion measurements which means that they might be misidentified. 

\subsection{Data sources for normal and millisecond radio pulsars}
\label{s:pulsar_sources}
For the normal radio pulsars we use the interferometric measurements of the parallax and proper motion based on several previous works: \cite{bbgt02,btgg03,ccl+01,ccv+04,cbv+09,dtbr09,dgb18,kvw+15}. To filter this list, we   select only pulsars with relative error in parallaxes and proper motions less than 0.25. All of the selected objects and their measurements are compiled and shown in Table~\ref{t:normal_psr}. In cases where  a measurement has some asymmetric uncertainties, we use the largest uncertainty value. In total our list contains 57 pulsars. A priori we do not expect to find any wide astrometric pairs because the spatial velocities of these pulsars exceed $40$~km~s$^{-1}$ (the velocity dispersion of NS progenitors is 10-15~km~s$^{-1}$) except for PSR J0614+2229 which has a velocity of $\approx 20$~km~s$^{-1}$.

For the MSPs we prepared the sample using the interferometric measurements from several previous studies:  \cite{cbv+09,dvt+08,dab+12,dbl+13,ysy+13} and collected the timing measurements from the following studies: \cite{abb+18,dcl+16,fwe+12,gsl+16,nbb14,rhc+16,slr14}. The vast majority of the MSP measurements are timing ones obtained from \cite{abb+18}. In our sample we kept only objects with relative error of parallax and proper motion of less than 0.25. The list of pulsars together with their measurements is shown in Table~\ref{t:msp}. It contains 21 MSPs, some of these objects have very small spatial velocities. In order to simplify our analysis, we use the proper motion measurements in the equatorial coordinate system, even though the measurements in the ecliptic system might be more precise. 

\subsection{Data sources for XRBs}
\label{s:xrb}
For our analysis we use two catalogues: \cite{liu_hmxb_2006} for HMXBs (114 entries) and \cite{liu_lmxb_2007} for LMXBs (187 entries) additionally checking details in the Simbad database \citep{simbad}. The Simbad database includes all objects from both of these catalogues and newer measurements.
%We noticed, that a few objects from the catalogues are not known to Simbad by their names. Here we list these sources together with their main names in the Simbad: XTE J1906+09 (XTE J1906+090), Swift J061223.0+701243.9 (SWIFT J061223.0+701243), GRS 1632-477 (INTREF 709), X1724-356 (INTREF 782), GRS 1741.2-2859 (Ginga 1741.6-2849) and GC X-2 ([CFS78] GC X-2).

To choose only Galactic XRBs, we make use of the hierarchy information from \texttt{h\_link} table in Simbad. We prepare our master list using ADQL query (\citealt{adql}, see Appendix~\ref{a:adql} for details) selecting objects with type HMXB or LMXB and measured magnitude in V filter. Additionally we go through the list of all objects which are mentioned in the catalogues but have different type in the Simbad database. There are 25 objects which were classified as HMXBs in \cite{liu_hmxb_2006}, but their type is different in the Simbad database. We found that these objects were most probably mis-classified before, e.g. V* V4641 Sgr mentioned as HMXB in \cite{liu_hmxb_2006} is classified as LMXB according to \cite{2014ApJ...784....2M} with the secondary mass $M_*=2.9\pm 0.4 M_\odot$; 2S 0053+604 appears to be Be star with an unusual X-ray emission properties  \citep{2002ApJ...575..435R}, V* RT Cru seems to be a symbiotic star according to \cite{2000A&AS..146..407B} and later works.

The cases of mis-classification or dubious classification are quite common for XRBs, because type identification requires long observations in multiple spectral ranges. We are interested in genuine XRBs with BH or NS companions, therefore we rely on the modern classification provided by the Simbad database. 

A list of HMXBs based on the Simbad database appeared to have one extragalactic object:  NGC 300 X-1 which was excluded manually. New XRBs were discovered since the publication of \cite{liu_hmxb_2006} and \cite{liu_lmxb_2007}, and therefore the total number of Galactic XRBs with known V magnitude is 165 (HMXBs) and 57 (LMXBs).  According to the Simbad database 44 HMXBs do not have V magnitude measurements. We tried to identify HMXBs based on their Gaia colors and absolute magnitude, but we were able to find only one system, CXOU J162046.2-513006 in the catalogue, see Appendix~\ref{a:identif}.

%\subsubsection{Filtering of the XRB list}

\subsubsection{Identification of the XRB sample in the Gaia dataset}

Our XRB identification is based on spatial coincidence and magnitude similarity. We had to choose both criteria, since in a small region of a few arcsec (comparable to the size of point spread function of e.g. \textit{XMM-Newton}), the second Gaia data release might contain up to tens of stars. The situation is especially difficult with the HMXBs which are typically found in star formation regions. 

The magnitude could provide a reliable additional information to facilitate the identification of HMXBs because their intrinsic colors and absolute magnitudes differ significantly from that of the background M/K stellar population. In the case of LMXBs the situation is more complicated, since their optical emission is dominated by the accretion disk.
The periodic optical variability and magnitude variation due to on/off state in the LMXBs could also have prevented us in some cases from correctly identifying them. In particular, in the case of V* V1727 Cyg the amplitude of the photometric variations reaches $\Delta B = 1.5$ mag. Our identification suggests that $D\approx 2$~kpc, a distance which is inconsistent with the claimed spectral type F of the stars according to \cite{1990AJ.....99..678C}.

 Practically, we searched for all stars in the Gaia database within a $15''$ cones around the XRBs locations.
 For six out of 165 HMXB and 13 out of 57 LMXBs there are no optical counterparts with similar magnitudes at the location provided by the catalogues or Simbad. 
 Our main test for the successful identification is a comparison of the $V$ magnitude with the Gaia $g$ magnitude. For objects with $B-V=0$ (A2V spectral type) the unabsorbed $g\approx V$ according to \cite{gaia_photometry}. For B2V (common HMXB optical counterpart) stars the expected unabsorbed $g-V\approx-0.05$, while for G5V stars $g-V=-0.25$. The interstellar reddening can significantly change this value, that is why we allow $|g-V| < 1.5$ as a confirmation of positive identification. We assume that a star at the smallest separation from the expected coordinate with the  $|g-V| < 1.5$ is the best counterpart for the XRB.

The six HMXBs which were not identified are: 
\begin{enumerate}
\item 4U 1954+31 ($g-V\approx -1.6$ expected -1.45 based on $B-V$, but within uncertainty of \citealt{gaia_photometry} equation), 
\item SXP 4.78 no candidates brighter than $g=17.33$ in $15''$ circle of its position although $B-V=0.1$ and $V=15.8$
\item EM* VRMF 55, the candidate with $g-V=-2$ is difficult to verify since $B$ is not measured
\item AX J0048.2-7309, the candidate with $g-V=2.13$ is difficult to verify since $B$ is not measured
\item KRL2007b 348, the color $B-V=1.62$ should correspond to $g-V\approx -0.97$ while in reality 1.64
\item INTEGRAL1 111, the color $B-V=2.2$ should correspond to $g-V\approx -1.63$ while actual $g-V\approx 1.8$
\end{enumerate}
The list of LMXBs with the magnitude identification problems looks quite similar and we therefore do not show it here. 

%Additionally we have excluded 2MASS J18453684+0051474 and RX J0100.2-7204 which are dim $V\approx 20$ and have no parallax measurement. 
From our compiled list of XRBs with the Gaia counterpart we selected only objects with relative errors in parallax $\sigma_\varpi /\varpi <0.25$ and proper motions $\sigma_\mu /\mu_\alpha <0.25$ and $\sigma_\mu /\mu_\delta <0.25$.
Additionally we checked the quality of the astrometric solution following \cite{2018A&A...616A...2L}.
%\begin{equation}
%\sqrt{\frac{\chi^2}{\nu'-5}} < 1.2\; \mathrm{max}\left(1, \exp(-0.2[G-19.5])\right) 
%\end{equation}
%where $\chi^2$ is \texttt{astrometric\_chi2\_al} and $\nu'$ is \texttt{astrometric\_n\_good\_obs\_al} from the Gaia database.
The final list consists of 28 HMXBs and 7 LMXBs, see Table~\ref{t:xrb_gaia}. 
%The reason for this filtering is that we are searching for wide astrometric pairs. Object with poorly measured parallax and proper motion will coincide much more often than ones with a good measurements. In general, we expect that the HMXBs follow the motion of young stellar population, so their parallaxes and proper motions are typical for their surroundings. 

\begin{table*}
   \centering
   \begin{tabular}{llrccrrrccc}
    \hline  
    HMXB  & Sp. type & Gaia counterpart & V     & g     &  $\varpi\pm \sigma_\varpi$ & $\mu_\alpha \pm \sigma_\alpha$ & $\mu_\delta \pm \sigma_\delta$\\
          & & Gaia DR2 ID      & (mag) & (mag) & (mag)  &  (mas / year) & (mas/year)\\
    \hline
HD 100199                 &  B0/1(III)n(e)  &  5333660129603575808  &  8.17  &  8.13  & $ 0.769  \pm  0.069 $ & $ -6.202  \pm  0.109 $ & $ 1.296  \pm  0.099 $  \\
GRO J2058+42              &  O9.5-B0IV-Ve  &  2065653598916388352  &  14.74  &  14.19  & $ 0.077  \pm  0.018 $ & $ -2.219  \pm  0.033 $ & $ -3.356  \pm  0.031 $  \\
LS 1698                   &  B0III/V:e  &  5352018121173519488  &  11.48  &  11.25  & $ 0.156  \pm  0.034 $ & $ -6.265  \pm  0.062 $ & $ 2.821  \pm  0.057 $  \\
UCAC2   4813819           &  B2III/B0V  &  5335021599905643264  &  13.4  &  14.75  & $ 0.236  \pm  0.027 $ & $ -3.464  \pm  0.039 $ & $ 1.18  \pm  0.04 $  \\
KRL2007b 84             &  B0e  &  5258414192353423360  &  15.27  &  13.9  & $ 0.244  \pm  0.033 $ & $ -4.844  \pm  0.064 $ & $ 3.546  \pm  0.055 $  \\
GSC 03588-00834           &  B0Ve  &  2162805896614571904  &  14.2  &  13.76  & $ 0.116  \pm  0.019 $ & $ -3.533  \pm  0.031 $ & $ -3.172  \pm  0.028 $  \\
HD  34921                 &  B0IVpe  &  184497471323752064  &  7.48  &  7.22  & $ 0.753  \pm  0.057 $ & $ 1.444  \pm  0.118 $ & $ -4.116  \pm  0.07 $  \\
AX J1700.2-4220  &  Be  &  5966213219190201856  &  9.15  &  8.68  & $ 0.616  \pm  0.063 $ & $ 1.214  \pm  0.123 $ & $ -1.42  \pm  0.096 $  \\
AX J1739.1-3020           &  O8.5Iab(f)  &  4056922100878037120  &  14.8  &  16.23  & $ 0.651  \pm  0.068 $ & $ 2.887  \pm  0.112 $ & $ 1.732  \pm  0.087 $  \\
BD+53  2790               &  O9.5Vep  &  2005653524280214400  &  9.84  &  9.74  & $ 0.267  \pm  0.03 $ & $ -4.256  \pm  0.058 $ & $ -3.11  \pm  0.048 $  \\
V* V572 Pup               &  B0.2IVe  &  5548261400354128768  &  12.74  &  12.48  & $ 0.103  \pm  0.024 $ & $ -1.575  \pm  0.037 $ & $ 2.149  \pm  0.045 $  \\
HD 249179                 &  B5ne  &  3431561565357225088  &  10.0  &  10.01  & $ 0.387  \pm  0.057 $ & $ 0.61  \pm  0.108 $ & $ -2.716  \pm  0.087 $  \\
V* GP Vel                 &  B0.5Ia  &  5620657678322625920  &  6.87  &  6.72  & $ 0.384  \pm  0.03 $ & $ -4.963  \pm  0.052 $ & $ 9.092  \pm  0.052 $  \\
SS 188                    &  OB  &  5541793213959987968  &  12.33  &  12.17  & $ 0.1  \pm  0.025 $ & $ -2.558  \pm  0.039 $ & $ 3.08  \pm  0.044 $  \\
V* V479 Sct               &  O(f)N6.5V  &  4104196427943626624  &  11.27  &  10.8  & $ 0.484  \pm  0.052 $ & $ 7.428  \pm  0.088 $ & $ -8.0  \pm  0.069 $  \\
BD+60    73               &  B1Ib  &  427234969757165952  &  9.66  &  9.45  & $ 0.274  \pm  0.031 $ & $ -1.801  \pm  0.043 $ & $ -0.366  \pm  0.039 $  \\
EM* AS   14               &  B2  &  414196617287885312  &  11.36  &  11.43  & $ 0.367  \pm  0.029 $ & $ -2.298  \pm  0.037 $ & $ -0.549  \pm  0.039 $  \\
4U 2238+60                &  Be  &  2201091578667140352  &  14.8  &  14.15  & $ 0.084  \pm  0.019 $ & $ -2.393  \pm  0.032 $ & $ -1.045  \pm  0.03 $  \\
IGR J01583+6713   &  B2IVe+  &  518990967445248256  &  14.43  &  13.7  & $ 0.098  \pm  0.018 $ & $ -1.144  \pm  0.024 $ & $ 0.298  \pm  0.03 $  \\
HD  74194                 &  O8.5Ib-II(f)p  &  5522306019626566528  &  7.55  &  7.44  & $ 0.424  \pm  0.03 $ & $ -7.453  \pm  0.056 $ & $ 5.805  \pm  0.048 $  \\
HD  63666                 &  B7IV/V  &  5489434710755238400  &  7.6  &  7.52  & $ 1.529  \pm  0.043 $ & $ -4.528  \pm  0.088 $ & $ 8.596  \pm  0.086 $  \\
WRAY 15-793               &  O9.5III/Ve  &  5336957010898124160  &  12.12  &  11.6  & $ 0.313  \pm  0.027 $ & $ -5.387  \pm  0.04 $ & $ 1.301  \pm  0.037 $  \\
HD  49798                 &  sdO6  &  5562023884304074240  &  8.287  &  8.21  & $ 1.969  \pm  0.064 $ & $ -4.11  \pm  0.113 $ & $ 5.669  \pm  0.118 $  \\
V* BP Cru                 &  B1.5Iaeq  &  6054569565614460800  &  10.66  &  9.76  & $ 0.253  \pm  0.035 $ & $ -5.303  \pm  0.051 $ & $ -2.166  \pm  0.049 $  \\
HD 245770                 &  O9/B0III/Ve  &  3441207615229815040  &  9.39  &  8.68  & $ 0.442  \pm  0.049 $ & $ -0.628  \pm  0.09 $ & $ -3.036  \pm  0.066 $  \\
IGR J17544-2619           &  O9Ib  &  4063908810076415872  &  12.94  &  11.67  & $ 0.352  \pm  0.05 $ & $ -0.65  \pm  0.084 $ & $ -0.534  \pm  0.066 $  \\
HD 226868                 &  O9.7Iabpvar  &  2059383668236814720  &  8.91  &  8.52  & $ 0.422  \pm  0.032 $ & $ -3.882  \pm  0.048 $ & $ -6.171  \pm  0.054 $  \\
CPD-63  2495              &  O9.5Ve  &  5862299960127967488  &  9.98  &  9.63  & $ 0.418  \pm  0.031 $ & $ -6.986  \pm  0.043 $ & $ -0.416  \pm  0.044 $  \\
    \hline
    LMXB \\
    \hline
V* V1727 Cyg             &    &  1978241050130301312  &  16.4  &  17.6  & $ 0.547  \pm  0.092 $ & $ -2.301  \pm  0.156 $ & $ -4.392  \pm  0.152 $ & \\
V* GU Mus                &  K3V-K7V  &  5234956524094035840  &  13.3  &  14.78  & $ 0.165  \pm  0.023 $ & $ -3.085  \pm  0.039 $ & $ 0.463  \pm  0.035 $ & \\
V* V934 Cen              &  M2III  &  6158386896580577152  &  7.85  &  6.91  & $ 1.851  \pm  0.063 $ & $ -39.583  \pm  0.103 $ & $ -6.028  \pm  0.144 $ & \\
V* V1333 Aql             &    &  4264296560926758144  &  14.8  &  14.2  & $ 0.186  \pm  0.043 $ & $ -5.009  \pm  0.089 $ & $ -6.141  \pm  0.069 $ & \\
V* HZ Her                &  B3/6ep  &  1338822021487330304  &  13.63  &  13.61  & $ 0.149  \pm  0.027 $ & $ -1.267  \pm  0.047 $ & $ -7.91  \pm  0.053 $ & \\
V* IL Lup                &  A2V  &  5986117923045613184  &  14.9  &  15.73  & $ 0.314  \pm  0.044 $ & $ -7.221  \pm  0.087 $ & $ -2.971  \pm  0.064 $ & \\
PSR J1023+0038           &  GV  &  3831382647922429952  &  17.31  &  16.27  & $ 0.728  \pm  0.143 $ & $ 4.751  \pm  0.135 $ & $ -17.348  \pm  0.135 $ & \\
    \hline
    \end{tabular}
    \caption{Catalogue of XRBs with their Gaia counterparts with well measured parallaxes and proper motions. }
    \label{t:xrb_gaia}
\end{table*}

\begin{figure*}
\begin{minipage}{0.48\linewidth}
	\includegraphics[width=\columnwidth]{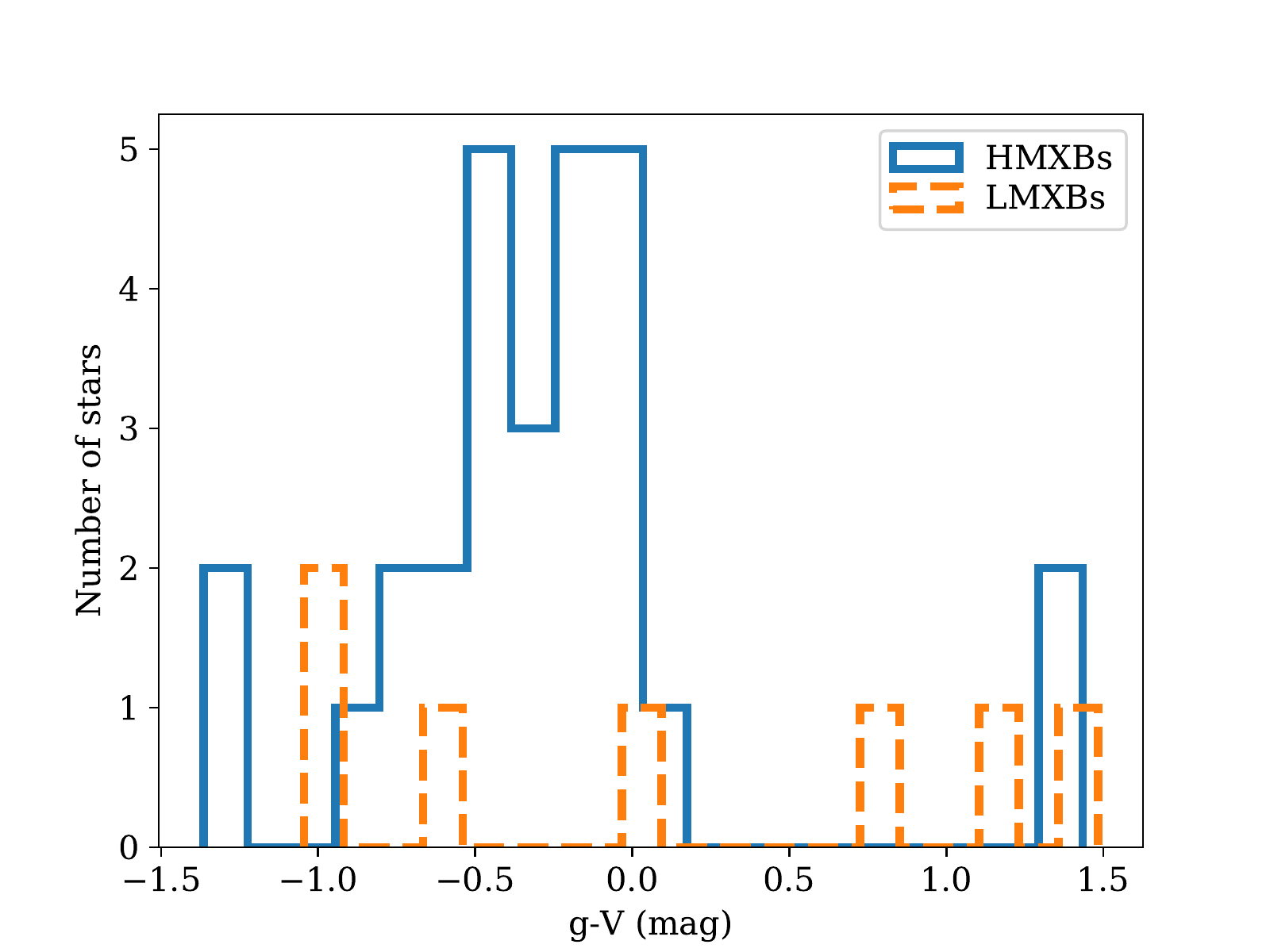}
\end{minipage}
\begin{minipage}{0.48\linewidth}
	\includegraphics[width=\columnwidth]{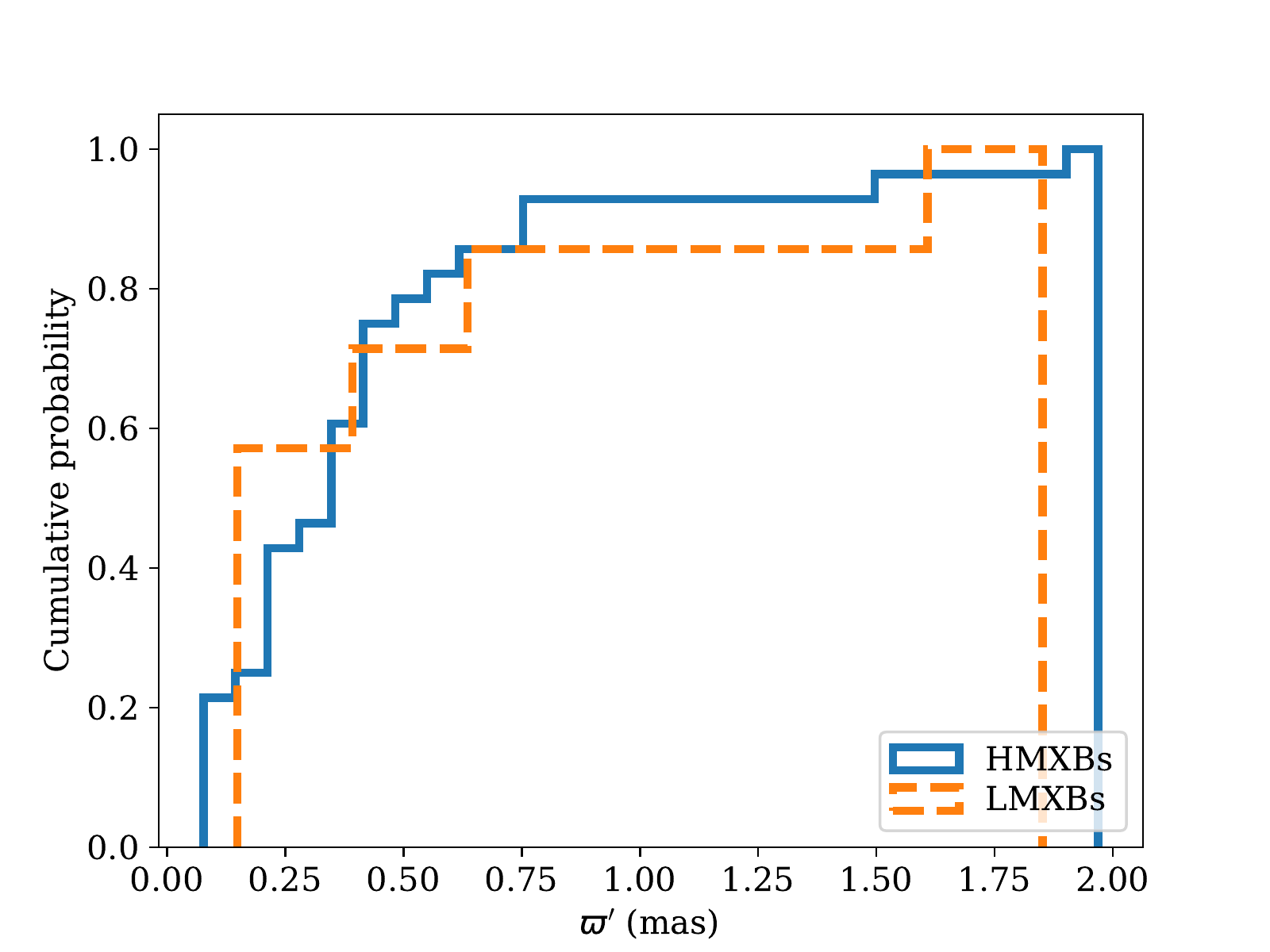}
\end{minipage}
\caption{Left panel: The distribution of observed colors for the XRBs matched with their Gaia counterparts. Right panel: The cumulative distribution of parallaxes for the Gaia counterparts of XRBs.}
    \label{f:hmxb_stat_properties}
\end{figure*}

We show the distribution of parallaxes and $g-V$ colors in Figure~\ref{f:hmxb_stat_properties}. The mean parallax for the 28 HMXBs is 0.44~mas and the mean $g$ magnitude is 10.78.  
There are two outliers with positive $g-V$ at this plot which is not expected. These two HMXBs are UCAC2 4813819 and AX J1739.1-3020, we believe these might be misidentified.

The mean parallax of seven LMXBs is 0.56~mas and the mean $g$ magnitude is 14.16. The main reason why we have such a limited statistics for LMXBs is that the second Gaia data release does not have precise astrometric solution for the majority of LMXBs, likely because LMXBs are faint and are strongly variable in the optical band. We find three outliers with unexpected color $g-V > 0$: V* V1727 Cyg, V* GU Mus and V* IL Lup.

% Among other factors, we need to mention, that
%It makes our search more challenging, since we need to filter out a contamination by 

\begin{figure*}
\begin{minipage}{0.48\linewidth}
	\includegraphics[width=\columnwidth]{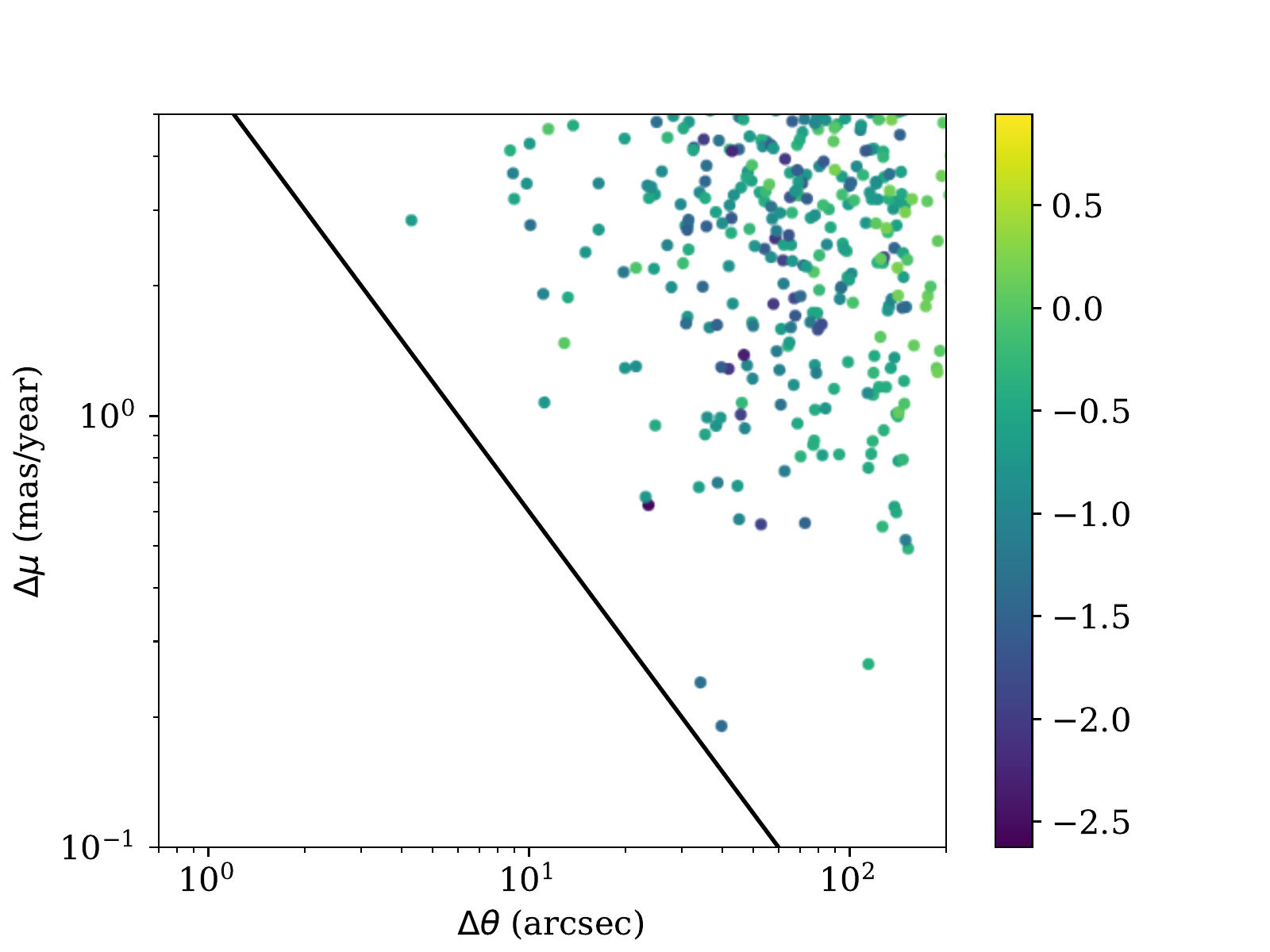}
\end{minipage}
\begin{minipage}{0.48\linewidth}
	\includegraphics[width=\columnwidth]{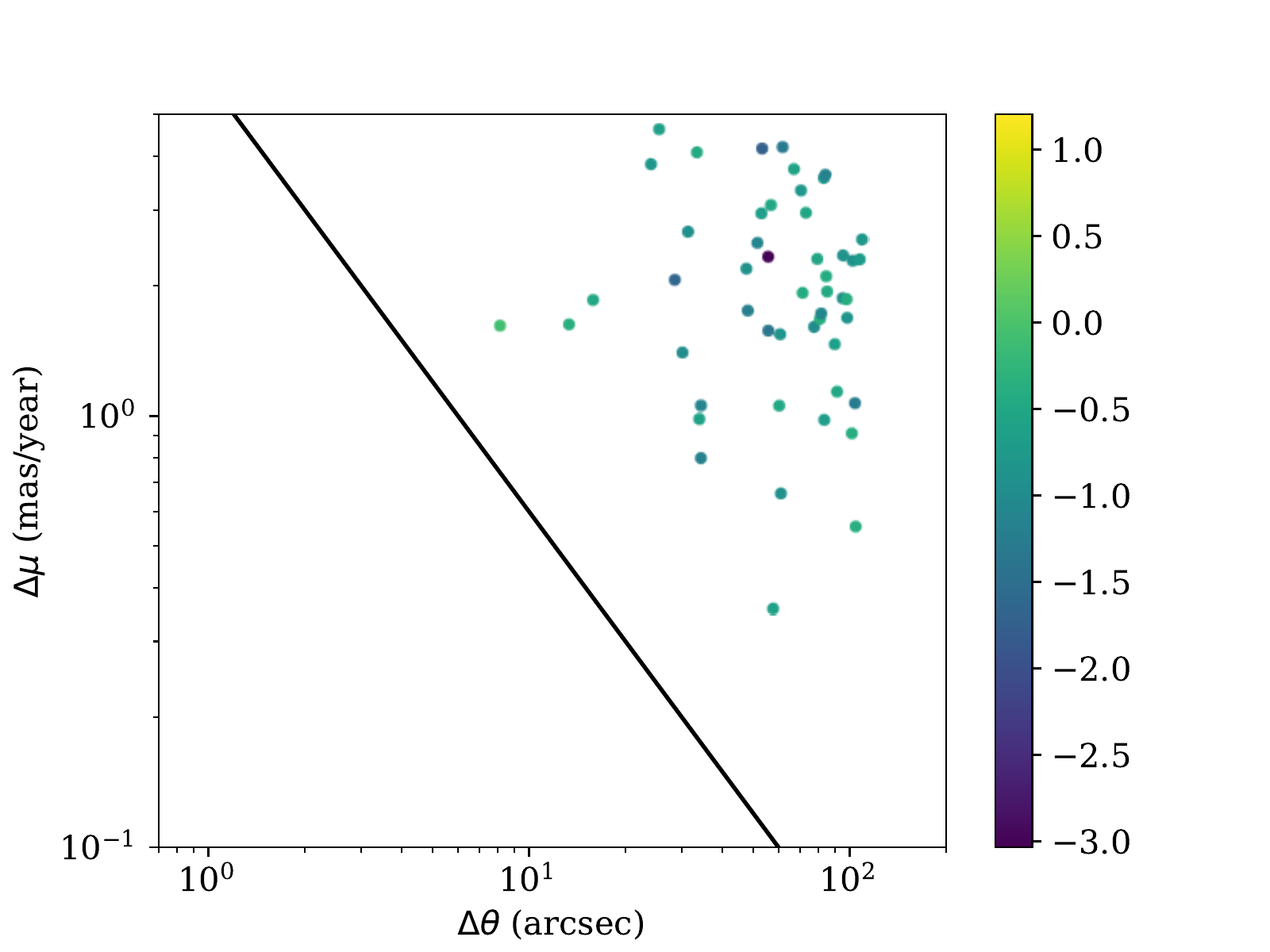}
\end{minipage}
\caption{The differences in proper motions vs. the angular separation for HMXBs (left panel) and LMXBs (right panel).  The color shows the logarithm of parallax difference.}
    \label{f:hmxb_separation}
\end{figure*}

%We deal with these difficulties as following. In the first case, we check if the Simbad database has coordinate for this particular XRB and proceed further with these new coordinates.  

%\subsection{Color-based identification of LMXBs}
%165 out of 210 HMXBs from the Simbad have measured V magnitudes. 57 out of 77 LMXBs %have measured V values.

\subsection{Results}
\begin{figure*}
\begin{minipage}{0.48\linewidth}
	\includegraphics[width=\columnwidth]{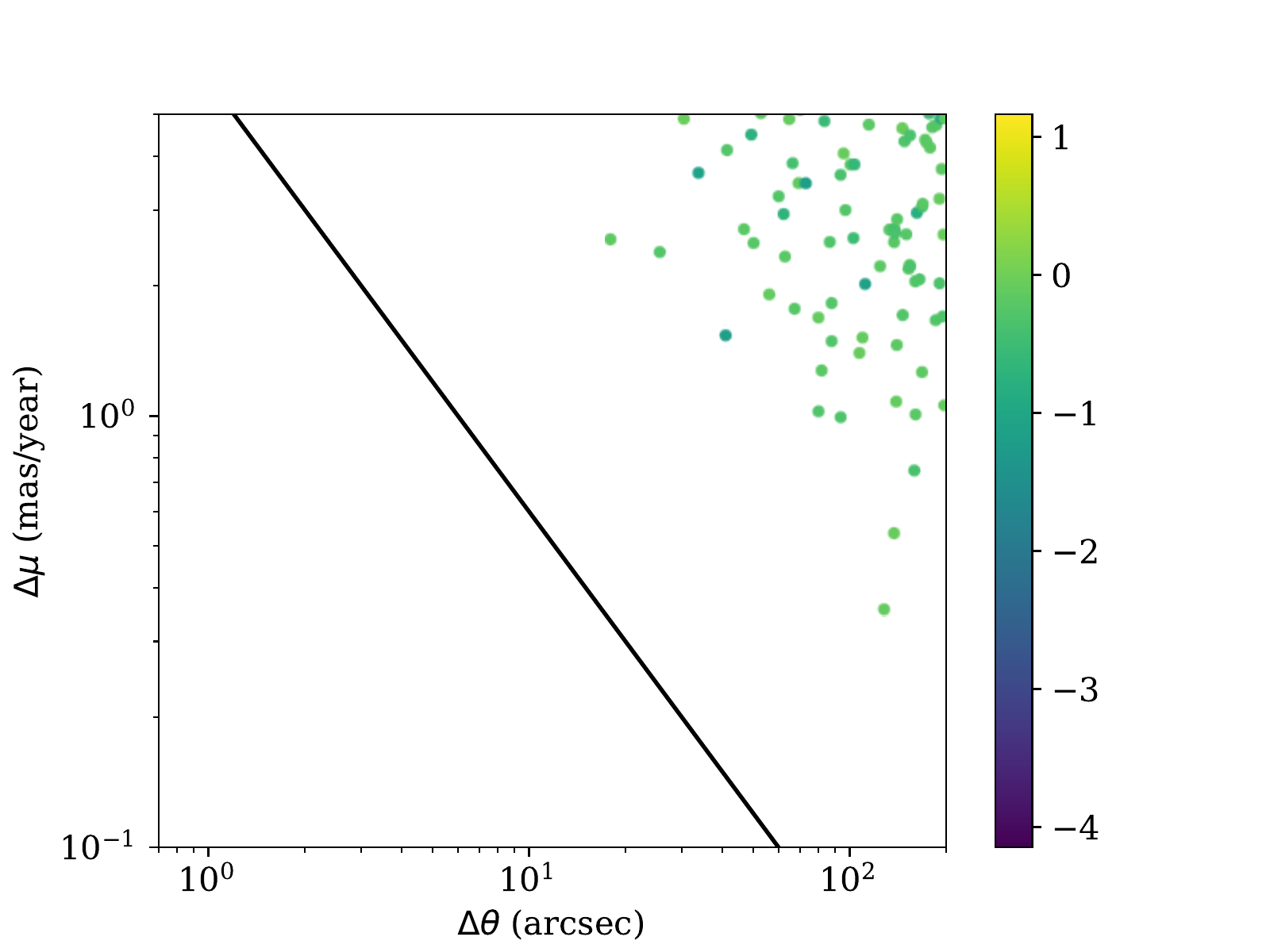}
\end{minipage}
\begin{minipage}{0.48\linewidth}
	\includegraphics[width=\columnwidth]{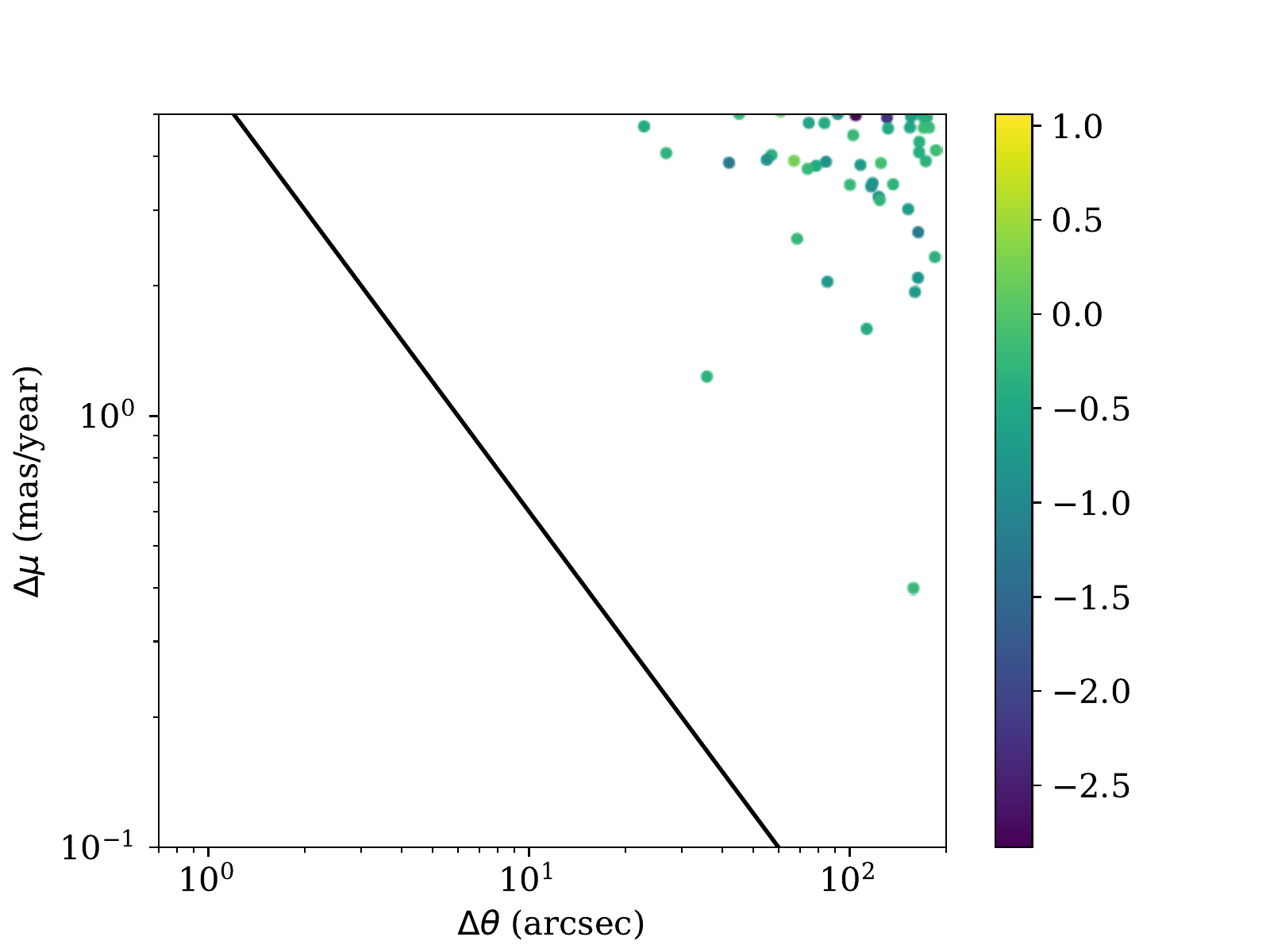}
\end{minipage}
\caption{The differences in the proper motions vs. the angular separation of PSRs (left panel) and MSPs (right panel). The color shows the logarithm of the parallax difference. }
    \label{f:psr_separation}
\end{figure*}

We found no reliable wide astrometric pairs of XRB or MSP or pulsar and a field star. 
We show the distribution of angular separations and differences in proper motions in Figures~\ref{f:hmxb_separation}~and~\ref{f:psr_separation}.
%We have found one pairing between the slowest pulsar in the sample PSR J0614+2229 and a field star and one pairing between HMXB SS118 and a field star. 
Below we analyze and discuss in more details the statistical significance of these results. 

\subsection{Constraints on fraction of weak natal kicks}

The direct comparison of the verification sample (O/B stars) with the non detection of reliable wide components of XRBs is not straightforward because some of the OB stars from the verification sample are isolated while all XRBs were originally binaries (we neglect here the dynamical formation in the field). To take this effect into account we have to rely on stellar statistics. 
\cite{moe2017} provides the following fraction of isolated, binary, triple and quadruple stars for primaries of $10~M_\odot$: 17.5\%, 35\%, 30\%, 17.5\%. If we neglect stellar interactions, progenitors of HMXBs were binaries in 42\%, triples in 36\% and quadruples in 22\% of cases. If none of the compact objects received a natal kick (which is known not to be the case) and the mass loss was adiabatic, such the third component could survive, we would have expected it that 58\% of HMXBs should have one or two bound companions. For a sample of early B stars the second, third or fourth component should have been identified in 83\% of the cases. This would suggest that a wide companion should  be identified 1.4 times more often in the sample of B stars than in sample of binary B stars.

Using the hypergeometric distribution we can compute how significant is it not to find any distant companions to HMXBs. We re-scale the fraction $0.044$ to $0.044 / 1.4 \approx 0.03$. For the moment let us assume that the total sample size is 165 objects (if we double the sample size it does not change the probability significantly). In this case the probability to find no distant companion is 0.39 and the probability to find a single distant component is 0.41. If we assume that only $10$~percent of NSs receive sufficiently small natal kicks to keep their wide companions then the probability to find no distant components becomes 0.92 and a probability to find one distant component is 0.08. 

At the moment the XRBs sample is to small to constrain the weakest natal kick through the suggested method. Future increase in the size of XRB sample in the Gaia will let us to constrain this hypothesis much better.

As for the normal radio pulsars, the probability to find no wide pairs among 57 pulsars is 0.07 (if the fraction of weak kicks is 100 per cent) which independently shows that most NSs receives strong natal kick. If the NSs receives weak natal kick in 10 per cent of the cases, non detection of ultra-wide pairs is compatible with expectations (0.83 probability to find none ultra-wide pair and 0.17 probability to find single ultra-wide pair).

As for millisecond radio pulsars the probabilities are: non detection has probability 0.5 (assuming 100 per cent fraction of weak kicks) and 0.93 (assuming 10 per cent fraction of weak kicks).

\section{Discussion and summary}

Using the second Gaia data release, we identified and characterized the population of ultra-wide massive OB binaries. We find that $4.4\pm0.5$ per cent of O-B5 stars have ultra-wide ($10^3-2\times10^5$ AU) companions down to the detection limit (G-stars).%, after accounting for false-positive identifications. 
%This is consistent with an ultra-wide multiplicity fraction of $48\pm6\%$ assuming the companion mass-function follows a Kroupa initial mass function down to the lowest stellar masses. 

In addition, we searched for astromentric wide companions to NS/BH binaries, by searching for third companions to HMXB, LMXB or MSP binaries or wide companions to normal radio pulsars. Since such wide binaries would be disrupted if the compact object received a natal kick larger than a few km s$^{-1}$, any finding of such wide binaries would provide a potential strong evidence for the existence of ultra-low or even zero-velocity natal kicks, while null finding could provide constraints on the fraction of systems that could receive such low-kicks. 

%Our search revealed no reliable pairs with angular separations less than $\Delta \theta < 25$~arcsec. We found one possible association between a low-velocity pulsar J0614+2229 and the star Gaia DR2 3376991158301978880. The possible wide companion has a projected orbital separation of $2.7\times 10^5$~a.u. However, this companion could be a chance alignment with a probability of $\approx 3-4$~per cent. If the association is real and the companion is physically bound to the neutron star it indicates that the neutron-star received an extremely low natal kick at birth, compatible with zero km~s$^{-1}$. A further study of the star and pulsar are necessary to confirm or reject this association.

%In the HMXBs sample, we find a possible association between SS118 and Gaia DR2 554179283600307712, again with a large orbital separation of $3.78\times 10^5$~a.u. for which the false positive probability is 2 percents. This companion can easily be confirmed or rejected using follow-up radial velocity measurements.

Finally, we note that the third Gaia data release will enable  us to probe the lowest NS and BH natal kicks using the same approach, but the better data will be sufficient to provide a much better statistically significant results and probe a larger sample of XRBs.

\section*{Acknowledgements}
A.I. thanks Evgeni Grishin and Yonadav Bary Ginat for useful discussions. A.I. and H.P. thanks anonymous referee for useful comments.

This work has made use of data from the European Space Agency (ESA) mission
{\it Gaia} (\url{https://www.cosmos.esa.int/gaia}), processed by the {\it Gaia}
Data Processing and Analysis Consortium (DPAC,
\url{https://www.cosmos.esa.int/web/gaia/dpac/consortium}). Funding for the DPAC
has been provided by national institutions, in particular the institutions
participating in the {\it Gaia} Multilateral Agreement.

This research has made use of the SIMBAD database,
operated at CDS, Strasbourg, France  

\bibliographystyle{mnras} 
\bibliography{xrb_bibl}

\appendix
\section{The lifetime of ultra-wide binaries}
\label{a:lifetime}
%Both wide potential component which we identify for HMXB and a normal pulsar have orbital separation of order of 1~pc. Moreover, we see a number of such extra-wide binary systems in our OB sample. Therefore, 
It is worth discussing whether ultra-wide binaries could potentially be bound for a long time.
According to \cite{binney_tremaine}  two type of objects play leading roles in the disruption of a wide binary: passing stars and molecular clouds.
For the passing stars we get the following estimate:
\begin{equation}
t_d \simeq 15\; \mathrm{Gyr} \frac{k_\mathrm{diff}}{0.002} \frac{10^4~\mathrm{a.u.}}{a} \frac{M_b}{2M_\odot}    
\end{equation}
where $M_b$ is the total mass of the binary, the diffusion coefficient is $k_\mathrm{diff} \simeq 0.022 / \ln \Lambda$ and $\ln \Lambda$ is:
\begin{equation}
\ln \Lambda \approx  \ln\left\{ \frac{\sigma_\mathrm{rel}^2a}{GM_p} \right\}= 13.5
\end{equation}
where we assume the dispersion of relative velocity to be $\sigma_\mathrm{rel}\approx 40$~km~s$^{-1}$ and mass of perturber  to be $M_p=1~M_\odot$ which means that a massive binary ($M_b=20~M_\odot$) with semi-majour axis of $2\times 10^5$~a.u. can stay bound up to $\approx 6$~Gyr. 
The effect of passing molecular clouds is much more severe:
\begin{equation}
t_d \simeq 380~\mathrm{Gyr} \left(\frac{10^4~\mathrm{a.u.}}{a}\right)^3 \frac{M_b}{2M_\odot}
\end{equation}
assuming similar values for the surface and volume density of giant molecular clouds as in \cite{binney_tremaine} which results in a timescale of 0.5~Gyr for massive binary with orbital separation of $2\times 10^5$. It is known that massive B stars, HMXBs and normal radio pulsars can live for $\sim 10^7$~years. Therefore, such extra wide binaries could indeed be found.
%6~Myr for the HMXB and 20~Myr for the pulsar. It is worth noting that the distribution of projected separations for massive OB stars decreases abruptly at $\approx 4\times 10^5$~a.u., see Figure~\ref{f:ver_hist} which corresponds to a disruption timescale of less than $10$~Myr which is comparable to the lifetime of the B-star.

%In \cite{1987ApJ...312..367W} the survival time for a wide binary with initial semi-major axis $a_0$ was estimated as:
%\begin{equation}
%t(a_0) = 1.8\times 10^4~\mathrm{Myr} \left(\frac{M}{M_\odot}\right) %\left(\frac{a_0}{0.1~\mathrm{pc}}\right)^{-1}   
%\end{equation}
%which gives around $10^{10}$~years for B2 star with mass $\approx 10M_\odot$ and orbital separation of $2$~pc. For a neutron star this value is one order of magnitude less. 
%It is known that massive B stars, HMXBs and normal radio pulsars can live for $\sim 10^7$~years. In particular, the spin-down age (ratio of period $P$ to period derivative $\dot P$) of PSR J0614+2229 is $\tau = {P/(2\dot P)} = 8.9\times 10^4$~years which is significantly less than 20~Myr.   
%Therefore, such extra wide binaries could indeed be found.

\section{ADQL requests}
\label{a:adql}
Here we show two ADQL requests for the Simbad database which helped us to form the initial HMXB lists before filtering:
\begin{verbatim}
SELECT main_id, ra, dec, V
FROM basic join allfluxes ON oid=oidref
WHERE otype = 'HMXB' and V is not Null 
and oid NOT IN (SELECT child FROM h_link)
\end{verbatim}
This is the request for the LMXB source list:
\begin{verbatim}
SELECT main_id, ra, dec, V
FROM basic join allfluxes ON oid=oidref
WHERE otype = 'LMXB' and V is not Null 
and oid NOT IN (SELECT child FROM h_link)
\end{verbatim}

We select massive stars (B0-B5) stars from the Simbdad database with following query:
\begin{verbatim}
SELECT main_id, ra, dec, V
FROM basic join allfluxes ON oid=oidref
WHERE ((sp_type like 'B0%') or (sp_type like 'B1%')  
or (sp_type like 'B2%') or  (sp_type like 'B3%') 
or (sp_type like 'B4%') or (sp_type like 'B5%'))
and V is not Null and V > 0 and plx_value > 0 
and plx_err > 0 and plx_value/ plx_err > 4 
and pm_err_maj > 0 and pmra / pm_err_maj > 4 
and pm_err_maj > 0 and  pmra / pm_err_maj > 4
and plx_value > 0.1 
and oid NOT IN (SELECT child FROM h_link)
order by main_id
\end{verbatim}

Below we show our ADQL request to select binary candidates to a particular massive star or remnant:
\begin{verbatim}
SELECT * 
FROM ( 
SELECT source_id,ra, dec, parallax, parallax_error, 
pmra, pmra_error, bp_rp, phot_g_mean_flux_over_error, 
phot_bp_mean_flux_over_error, parallax_over_error, 
pmdec, pmdec_error, phot_g_mean_mag, 
astrometric_chi2_al, astrometric_n_good_obs_al, 
DISTANCE(POINT('ICRS',ra,dec),
POINT('ICRS', RA, DEC)) AS dist, 
abs(parallax - PAR) as dvarpi, 
sqrt(power(parallax_error, 2) + PAR_ERR) 
as sigma_delta_varpi, 
sqrt(power(pmra - PRMT_RA, 2) + 
power(pmdec - PRMT_DEC, 2))
as delta_mu, pmra - PRMT_RA as delta_mura, 
pmdec - PRMT_DEC as delta_mudec, 
1.0 / (sqrt(power(pmra - PRMT_RA, 2) +
power(pmdec - PRMT_DEC, 2))) *
sqrt( (power(pmra_error,2) + power(PRMT_RA_ERR, 2)) * 
power(pmra - PRMT_RA, 2) + 
(power(pmdec_error, 2.0) + power(PRMT_DEC_ERR,2)) 
* power(pmdec - PRMT_DEC, 2.0)) as sigma_delta_mu, 
0.53 * power (parallax, 1.5) 
/ sqrt(DISTANCE(POINT('ICRS',ra,dec), 
POINT('ICRS', RA , DEC)) * 3600) as mu_orb 
FROM gaiadr2.gaia_source 
WHERE  abs(parallax) / parallax_error > 3 
and parallax > 0 
and abs(pmra) / pmra_error > 3 and 
sqrt(astrometric_chi2_al / 
(astrometric_n_good_obs_al - 5)) 
< 1.2 *  0.5 *
((exp(-0.2 * (phot_g_mean_mag - 19.5)) + 1.0) 
+ abs(exp(-0.2 * (phot_g_mean_mag - 19.5)) - 1.0)) 
and abs(pmdec) / pmdec_error > 3 ) as t 
WHERE dvarpi < 2 * sigma_delta_varpi 
and delta_mu < (mu_orb + 2 * sigma_delta_mu)  
and sigma_delta_varpi < 0.12 and delta_mu <  3 * mu_orb 
and sigma_delta_mu < 0.16 
and 1=CONTAINS(POINT('ICRS',ra,dec), 
CIRCLE('ICRS',RA , DEC , ANGLE )) 
ORDER by dist
\end{verbatim}
where RA, DEC, PAR, PAR\_ERR, PRMT\_RA, PRMT\_DEC, PRMT\_RA\_ERR, PRMT\_DEC\_ERR are measured values for the primary star, namely: right ascension, declination, parallax, parallax error, proper motion in the right ascension direction and declination direction as well as their errors. ANGLE is the size of the requested area.

\section{Identification of HMXBs based on color}
\label{a:identif}
In order to increase statistical significance we tried to increase the sample of identified HXMBs based on color properties. To do so, we selected all Galactic HMXBs with no measured V magnitude:
\begin{verbatim}
SELECT main_id, ra, dec, V
FROM basic join allfluxes ON oid=oidref
WHERE otype = 'HMXB' and V is Null 
and oid NOT IN (SELECT child FROM h_link)
\end{verbatim}
After this, we checked colors and absolute magnitude (with no correction for reddening) of all stars in the second Gaia data release at angular separation of $15''$. We choose as potential candidates only stars with $-0.5<B_p - R_p<2$ and $-5<g + 15 \log_{10} \varpi + 5.0<3$. These boundaries were found using the scattering of colors and absolute magnitudes of stars from our verification sample.
Initial selection identified 13 candidates for 44 HMXBs and only one of them has good enough measurements of parallax and proper motion.

\section{List of normal radio pulsars}

\begin{table*}
   \centering
   \begin{tabular}{lrrrccccccc}
    \hline  
    PSR name  &  $\varpi\pm \sigma_\varpi$ & $\mu_\alpha \pm \sigma_\alpha$ & $\mu_\delta \pm \sigma_\delta$ & Ref. \\
          &  (mas / year) & (mas/year)\\
    \hline
J0034-0721  & $ 0.93  \pm  0.08 $ & $ 10.37  \pm  0.08 $ & $ -11.13  \pm  0.16 $ &  7  \\
J0055+5117  & $ 0.349  \pm  0.055 $ & $ 10.49  \pm  0.085 $ & $ -17.352  \pm  0.204 $ &  1  \\
J0102+6537  & $ 0.399  \pm  0.045 $ & $ 9.252  \pm  0.081 $ & $ 1.828  \pm  0.206 $ &  1  \\
J0108+6608  & $ 0.468  \pm  0.035 $ & $ -32.754  \pm  0.036 $ & $ 35.162  \pm  0.051 $ &  1  \\
J0139+5814  & $ 0.37  \pm  0.04 $ & $ -19.11  \pm  0.07 $ & $ -16.6  \pm  0.07 $ &  7  \\
J0147+5922  & $ 0.495  \pm  0.093 $ & $ -6.38  \pm  0.101 $ & $ 3.826  \pm  0.097 $ &  1  \\
J0157+6212  & $ 0.554  \pm  0.039 $ & $ 1.521  \pm  0.105 $ & $ 44.811  \pm  0.048 $ &  1  \\
J0323+3944  & $ 1.051  \pm  0.04 $ & $ 26.484  \pm  0.059 $ & $ -30.78  \pm  0.029 $ &  1  \\
J0332+5434  & $ 0.595  \pm  0.025 $ & $ 16.969  \pm  0.029 $ & $ -10.379  \pm  0.058 $ &  1  \\
J0358+5413  & $ 0.91  \pm  0.16 $ & $ 9.2  \pm  0.18 $ & $ 8.17  \pm  0.39 $ &  5  \\
J0454+5543  & $ 0.84  \pm  0.05 $ & $ 53.34  \pm  0.06 $ & $ -17.56  \pm  0.14 $ &  7  \\
J0601-0527  & $ 0.478  \pm  0.045 $ & $ -7.348  \pm  0.077 $ & $ -15.227  \pm  0.105 $ &  1  \\
J0614+2229  & $ 0.282  \pm  0.031 $ & $ -0.233  \pm  0.053 $ & $ -1.224  \pm  0.065 $ &  1  \\
J0629+2415  & $ 0.333  \pm  0.054 $ & $ 3.629  \pm  0.193 $ & $ -4.607  \pm  0.153 $ &  1  \\
J0630-2834  & $ 3.009  \pm  0.409 $ & $ -46.3  \pm  0.99 $ & $ 21.26  \pm  0.52 $ &  6  \\
J0659+1414  & $ 3.47  \pm  0.36 $ & $ 44.07  \pm  0.63 $ & $ -2.4  \pm  0.29 $ &  8  \\
J0729-1836  & $ 0.489  \pm  0.098 $ & $ -13.072  \pm  0.125 $ & $ 13.252  \pm  0.456 $ &  1  \\
J0814+7429  & $ 2.31  \pm  0.04 $ & $ 24.02  \pm  0.09 $ & $ -43.96  \pm  0.35 $ &  2  \\
J0820-1350  & $ 0.51  \pm  0.04 $ & $ 21.64  \pm  0.09 $ & $ -39.44  \pm  0.05 $ &  7  \\
J0826+2637  & $ 2.01  \pm  0.013 $ & $ 62.994  \pm  0.021 $ & $ -96.733  \pm  0.085 $ &  1  \\
J0922+0638  & $ 0.83  \pm  0.13 $ & $ 18.8  \pm  0.9 $ & $ 86.4  \pm  0.7 $ &  4,9  \\
J0953+0755  & $ 3.82  \pm  0.07 $ & $ -2.09  \pm  0.08 $ & $ 29.46  \pm  0.07 $ &  2  \\
J1136+1551  & $ 2.687  \pm  0.018 $ & $ -73.785  \pm  0.031 $ & $ 366.569  \pm  0.072 $ &  1  \\
J1239+2453  & $ 1.16  \pm  0.08 $ & $ -106.82  \pm  0.17 $ & $ 49.92  \pm  0.18 $ &  2  \\
J1321+8323  & $ 0.968  \pm  0.14 $ & $ -52.674  \pm  0.099 $ & $ 32.373  \pm  0.204 $ &  1  \\
J1509+5531  & $ 0.47  \pm  0.03 $ & $ -73.64  \pm  0.05 $ & $ -62.65  \pm  0.09 $ &  7  \\
J1532+2745  & $ 0.624  \pm  0.096 $ & $ 1.542  \pm  0.127 $ & $ 18.932  \pm  0.118 $ &  1  \\
J1543+0929  & $ 0.13  \pm  0.02 $ & $ -7.61  \pm  0.06 $ & $ -2.87  \pm  0.07 $ &  7  \\
J1559-4438  & $ 0.384  \pm  0.081 $ & $ 1.52  \pm  0.14 $ & $ 13.15  \pm  0.05 $ &  6  \\
J1607-0032  & $ 0.91  \pm  0.046 $ & $ -26.501  \pm  0.066 $ & $ 27.082  \pm  0.451 $ &  1  \\
J1623-0908  & $ 0.586  \pm  0.101 $ & $ -10.769  \pm  0.131 $ & $ 23.509  \pm  0.166 $ &  1  \\
J1645-0317  & $ 0.252  \pm  0.028 $ & $ -1.011  \pm  0.051 $ & $ 20.523  \pm  0.205 $ &  1  \\
J1703-1846  & $ 0.348  \pm  0.049 $ & $ -0.751  \pm  0.102 $ & $ 16.962  \pm  0.23 $ &  1  \\
J1754+5201  & $ 0.16  \pm  0.029 $ & $ -3.95  \pm  0.047 $ & $ 1.101  \pm  0.072 $ &  1  \\
J1820-0427  & $ 0.351  \pm  0.055 $ & $ -7.318  \pm  0.074 $ & $ 15.883  \pm  0.088 $ &  1  \\
J1833-0338  & $ 0.408  \pm  0.067 $ & $ -17.409  \pm  0.158 $ & $ 15.038  \pm  0.337 $ &  1  \\
J1840+5640  & $ 0.657  \pm  0.065 $ & $ -31.212  \pm  0.033 $ & $ -29.079  \pm  0.082 $ &  1  \\
J1901-0906  & $ 0.51  \pm  0.067 $ & $ -7.531  \pm  0.045 $ & $ -18.211  \pm  0.159 $ &  1  \\
J1913+1400  & $ 0.185  \pm  0.027 $ & $ -5.265  \pm  0.072 $ & $ -8.927  \pm  0.065 $ &  1  \\
J1932+1059  & $ 2.78  \pm  0.06 $ & $ 94.06  \pm  0.09 $ & $ 43.24  \pm  0.17 $ &  3  \\
J1937+2544  & $ 0.318  \pm  0.031 $ & $ -10.049  \pm  0.042 $ & $ -13.055  \pm  0.039 $ &  1  \\
J2006-0807  & $ 0.424  \pm  0.101 $ & $ -6.176  \pm  0.07 $ & $ -10.616  \pm  0.174 $ &  1  \\
J2018+2839  & $ 1.03  \pm  0.1 $ & $ -2.64  \pm  0.21 $ & $ -6.17  \pm  0.38 $ &  2  \\
J2022+2854  & $ 0.61  \pm  0.08 $ & $ -3.46  \pm  0.17 $ & $ -23.73  \pm  0.21 $ &  3  \\
J2022+5154  & $ 0.78  \pm  0.07 $ & $ -5.03  \pm  0.27 $ & $ 10.96  \pm  0.17 $ &  3  \\
J2048-1616  & $ 1.05  \pm  0.03 $ & $ 113.16  \pm  0.02 $ & $ -4.6  \pm  0.28 $ &  7  \\
J2055+3630  & $ 0.17  \pm  0.03 $ & $ 1.04  \pm  0.04 $ & $ -2.46  \pm  0.13 $ &  7  \\
J2113+2754  & $ 0.704  \pm  0.023 $ & $ -27.981  \pm  0.052 $ & $ -54.432  \pm  0.096 $ &  1  \\
J2113+4644  & $ 0.454  \pm  0.077 $ & $ 9.525  \pm  0.148 $ & $ 8.846  \pm  0.09 $ &  1  \\
J2144-3933  & $ 6.051  \pm  0.56 $ & $ -57.89  \pm  0.88 $ & $ -155.9  \pm  0.54 $ &  6  \\
J2149+6329  & $ 0.356  \pm  0.072 $ & $ 15.786  \pm  0.131 $ & $ 11.255  \pm  0.284 $ &  1  \\
J2157+4017  & $ 0.28  \pm  0.06 $ & $ 16.13  \pm  0.1 $ & $ 4.12  \pm  0.12 $ &  7  \\
J2225+6535  & $ 1.203  \pm  0.204 $ & $ 147.22  \pm  0.243 $ & $ 126.532  \pm  0.115 $ &  1  \\
J2305+3100  & $ 0.223  \pm  0.033 $ & $ -3.737  \pm  0.082 $ & $ -15.571  \pm  0.163 $ &  1  \\
J2313+4253  & $ 0.93  \pm  0.07 $ & $ 24.15  \pm  0.1 $ & $ 5.95  \pm  0.13 $ &  7  \\
J2346-0609  & $ 0.275  \pm  0.036 $ & $ 37.39  \pm  0.042 $ & $ -20.23  \pm  0.107 $ &  1  \\
J2354+6155  & $ 0.412  \pm  0.043 $ & $ 22.755  \pm  0.056 $ & $ 4.888  \pm  0.033 $ &  1  \\
    \hline
    \end{tabular}
    \caption{List of normal radio pulsars used in the search.  1 -- \protect\cite{dgb18}, 2 -- \protect\cite{bbgt02}, 3 -- \protect\cite{kvw+15}, 4 -- \protect\cite{ccl+01}, 5 -- \protect\cite{ccv+04}, 6 -- \protect\cite{dtbr09}, 7 -- \protect\cite{cbv+09}, 8 -- \protect\cite{btgg03}, 9 -- \protect\cite{2003AJ....126.3090B} }
    \label{t:normal_psr}
\end{table*}

\section{List of MSPs}

\begin{table*}
   \centering
   \begin{tabular}{lrrrccccccc}
    \hline  
    PSR name  &  $\varpi\pm \sigma_\varpi$ & $\mu_\alpha \pm \sigma_\alpha$ & $\mu_\delta \pm \sigma_\delta$ & Ref. \\
          &  (mas / year) & (mas/year)\\
    \hline
J0437-4715  & $ 6.396  \pm  0.054 $ & $ 121.679  \pm  0.052 $ & $ -71.82  \pm  0.086 $ &  1  \\
J0613-0200  & $ 0.9  \pm  0.2 $ & $ 1.85  \pm  0.02 $ & $ -10.35  \pm  0.04 $ &  2  \\
J0751+1807  & $ 0.82  \pm  0.17 $ & $ -2.73  \pm  0.05 $ & $ -13.4  \pm  0.3 $ &  3  \\
J1012+5307  & $ 0.71  \pm  0.17 $ & $ 2.609  \pm  0.008 $ & $ -25.482  \pm  0.011 $ &  3  \\
J1022+1001  & $ 1.387  \pm  0.041 $ & $ -14.921  \pm  0.05 $ & $ 5.611  \pm  0.035 $ &  4  \\
J1023+0038  & $ 0.731  \pm  0.022 $ & $ 4.76  \pm  0.03 $ & $ -17.34  \pm  0.04 $ &  5  \\
J1300+1240  & $ 1.41  \pm  0.08 $ & $ 46.44  \pm  0.08 $ & $ -84.87  \pm  0.32 $ &  6  \\
J1455-3330  & $ 0.99  \pm  0.22 $ & $ 7.88  \pm  0.05 $ & $ -1.9  \pm  0.12 $ &  7  \\
J1600-3053  & $ 0.5  \pm  0.07 $ & $ -0.98  \pm  0.02 $ & $ -7.1  \pm  0.06 $ &  2  \\
J1614-2230  & $ 1.5  \pm  0.1 $ & $ 3.8  \pm  0.01 $ & $ -32.5  \pm  0.7 $ &  2  \\
J1643-1224  & $ 1.27  \pm  0.19 $ & $ 5.94  \pm  0.05 $ & $ 3.94  \pm  0.18 $ &  8  \\
J1713+0747  & $ 0.95  \pm  0.06 $ & $ 4.75  \pm  0.17 $ & $ -3.67  \pm  0.16 $ &  9  \\
J1738+0333  & $ 0.68  \pm  0.05 $ & $ 7.037  \pm  0.005 $ & $ 5.073  \pm  0.012 $ &  10  \\
J1741+1351  & $ 0.6  \pm  0.1 $ & $ -8.98  \pm  0.02 $ & $ -7.42  \pm  0.02 $ &  2  \\
J1744-1134  & $ 2.3  \pm  0.1 $ & $ 18.8  \pm  0.01 $ & $ -9.29  \pm  0.06 $ &  2  \\
J1909-3744  & $ 0.92  \pm  0.03 $ & $ -9.516  \pm  0.004 $ & $ -35.77  \pm  0.01 $ &  2  \\
J1918-0642  & $ 0.9  \pm  0.1 $ & $ -7.15  \pm  0.02 $ & $ -5.97  \pm  0.05 $ &  2  \\
J2043+1711  & $ 0.64  \pm  0.08 $ & $ -5.72  \pm  0.01 $ & $ -10.84  \pm  0.02 $ &  2  \\
J2124-3358  & $ 2.4  \pm  0.4 $ & $ -14.14  \pm  0.04 $ & $ -50.08  \pm  0.09 $ &  8  \\
J2145-0750  & $ 1.603  \pm  0.063 $ & $ -9.491  \pm  0.052 $ & $ -9.114  \pm  0.09 $ &  4  \\
J2222-0137  & $ 3.743  \pm  0.01 $ & $ 44.72  \pm  0.02 $ & $ -5.64  \pm  0.06 $ &  11  \\
    \hline
    \end{tabular}
    \caption{List of MSPs used in the search.  1 -- \protect\cite{dvt+08}, 2 -- \protect\cite{abb+18}, 3 -- \protect\cite{dcl+16}, 4 -- \protect\cite{dgb18}, 5 -- \protect\cite{dab+12}, 6 -- \protect\cite{ysy+13}, 7 -- \protect\cite{gsl+16}, 8 -- \protect\cite{rhc+16}, 9 -- \protect\cite{cbv+09}, 10 -- \protect\cite{fwe+12}, 11 -- \protect\cite{dbl+13},  }
    \label{t:msp}
\end{table*}

\begin{table*}
   \centering
   \begin{tabular}{lrccccccccc}
    \hline  
    MSP  & Gaia counterpart &  g     &  $\varpi\pm \sigma_\varpi$ & $\mu_\alpha \pm \sigma_\alpha$ & $\mu_\delta \pm \sigma_\delta$\\
         & Gaia DR2 ID      &  (mag) & (mag)  &  (mas / year) & (mas/year)\\
    \hline
J0437-4715  &  4789864076732331648  &  20.41  & $ 8.325  \pm  0.678 $ & $ 122.864  \pm  1.197 $ & $ -71.166  \pm  1.67 $  \\
J1012+5307  &  851610861391010944  &  19.63  & $ 1.326  \pm  0.414 $ & $ 2.977  \pm  0.526 $ & $ -26.944  \pm  0.632 $  \\
J1023+0038  &  3831382647922429952  &  16.27  & $ 0.728  \pm  0.143 $ & $ 4.751  \pm  0.135 $ & $ -17.348  \pm  0.135 $  \\
J1024-0719  &  3775277872387310208  &  19.18  & $ 0.529  \pm  0.426 $ & $ -35.519  \pm  0.64 $ & $ -47.932  \pm  0.659 $  \\
J1843-1113  &  4106823440438736384  &  19.8  & $ 0.787  \pm  0.643 $ & $ -3.241  \pm  1.83 $ & $ -7.274  \pm  2.171 $  \\
    \hline
    \end{tabular}
    \caption{Identified optical counterparts to millisecond radio pulsars.}
    \label{t:ident_gaia}
\end{table*}

\section{Formation of ultra-wide binaries due to a strong natal kick}
\label{s:population}
The strong natal kick could lead to a formation of ultra wide binary if it has preferable orientation and magnitude. For example, if the binary escape velocity is $100$~km~s$^{-1}$ and the natal kick of neutron star is 99~km~s$^{-1}$ (assuming zero mass loss for a moment), it would lead to a formation of a binary with large semi-major axis and eccentricity close to 1. However, this happens quite infrequently. To check the likelihood for such formation channel we performed a binary stellar population synthesis using the code SeBa \citep{1996A&A...309..179P,2012A&A...546A..70T,2013A&A...557A..87T}.    

In the modeling we mostly use the standard assumptions about the initial distribution of semi-major axis, eccentricities, initial masses and mass fractions, see e.g. \cite{2018A&A...619A..53T}. In particular we choose to model the distribution of semi-major axis following a flat distribution in logarithm of semi-major axis ranging from 0 till $1000$~a.u. \citep{1983ARA&A..21..343A}. The smaller range is chosen to exclude initially wide systems which survived a weak natal kick.  For primary star the mass ranges between 4 and 25 solar masses. The mass is chosen from  the Kroupa mass function \citep{2001MNRAS.322..231K}. The secondary masses are drawn from flat distribution $0<M_2/M_1 <1$  \citep{Duc+13}. The initial eccentricities follow the thermal distribution \citep{1975MNRAS.173..729H}. 

We compute the evolution of stars for 60~Myr because we are primary interested in normal radio pulsars with ultra-wide companion. The lifetime of normal radio pulsar seems to be comparable to $10$~Myr, the lifetime of lightest star which forms a neutron star in our simulations is $\approx45$~Myr.
We choose to model two very different natal kick velocity distributions, namely the \cite{verbunt2017} (model A) and \cite{hobbs2005} (model B). The former one has significantly large fraction of weak natal kicks, while the later one has small number of weak natal kicks.

A fraction of close binaries goes through the common envelope evolution before or after the primary supernova explosion which imparts the natal kick. The common envelope evolution is modelled through so called $\alpha$-prescription \citep{1976IAUS...73...75P,1987A&A...183...47D,1988ApJ...329..764L,1990ApJ...358..189D} with $\alpha\lambda=2$ (model 2) The systems which went through common envelope after formation of neutron star are not interesting for us, because these are systems with extremely short orbital periods. The binaries which went through an episode of the common envelope evolution before the primary supernova explosion might have different orbital separation depending on parameters chosen for the common envelope. Therefore we decide to vary this parameter too, in our alternative model we choose $\alpha\lambda=0.5$ (model 05). For each model we compute a sample starting from 5 million initial binaries.

Using the results of the evolution we select only bound detached systems where only one of stars is a neutron star at the end of evolution. We also checked the mass of the secondary and choose only systems where secondary is more massive than $0.5M_\odot$. The results of our simulations are presented in Table~\ref{t:pop_synthesis}. 

The largest apastron distance which we were able to find is $2.9\times 10^4$a.u for A2 model. A fraction of systems with large apoastron distances is $\approx 10^{-4}$ of all NS population.
%Only 70 out of 752432 stars (fraction $10^{-4}$) have apoastron distance larger than $10^3$~a.u.  }

\begin{table}
    \centering
    \begin{tabular}{ccccc}
    \hline
    Model & max $A_a$ & Fraction of $A_a>10^3$~a.u. \\
          &  a.u.                \\
    \hline
    A2    & $2.9\times 10^4$ & $9\times 10^{-5}$  \\
    A05   & $1.1\times 10^4$ & $9\times 10^{-5}$ \\
    B05   &  $4.2\times 10^3$ & $2\times 10^{-5}$ \\
    B2    & $2.3\times 10^4$  & $6\times 10^{-6}$ \\
         \hline
    \end{tabular}
    \caption{Results of the binary stellar population synthesis. max $A_a$ is the largest apoastron distance found in the sample. Fraction $A_a>10^3$~a.u. is the fraction of bound systems with an NS and secondary star with mass more than $0.5M_\odot$ relative to number of all formed neutron stars. }
    \label{t:pop_synthesis}
\end{table}

%The Acknowledgements section is not numbered. Here you can thank helpful
%colleagues, acknowledge funding agencies, telescopes and facilities used etc.

%%%%%%%%%%%%%%%%%%%%%%%%%%%%%%%%%%%%%%%%%%%%%%%%%%



%\bibliographystyle{mnras}

%\begin{thebibliography}{99}
%\end{thebibliography}

%%%%%%%%%%%%%%%%%%%%%%%%%%%%%%%%%%%%%%%%%%%%%%%%%%

%%%%%%%%%%%%%%%%% APPENDICES %%%%%%%%%%%%%%%%%%%%%

%\appendix



%%%%%%%%%%%%%%%%%%%%%%%%%%%%%%%%%%%%%%%%%%%%%%%%%%

% Don't change these lines
\bsp	% typesetting comment
\label{lastpage}
\end{document}